%% file: apssamp.tex
\documentclass[%
 reprint,
superscriptaddress,
 amsmath,amssymb,
 aps,
]{revtex4-2}

\usepackage{graphicx}
\usepackage{dcolumn}
\usepackage{float} 
\usepackage{bm}
\usepackage{bbm}
\usepackage{xcolor}

\renewcommand{\selectlanguage}[1]{}

\begin{document}
\preprint{APS/123-QED}

\title{\textbf{Renormalization group of topological scattering networks}}

\author{Zhe Zhang}
\affiliation{Laboratory of Wave Engineering, \'Ecole Polytechnique F\'ed\'erale de Lausanne (EPFL), CH-1015 Lausanne, Switzerland}

\author{Yifei Guan}
\affiliation{Institute of Physics, \'Ecole Polytechnique F\'ed\'erale de Lausanne (EPFL), CH-1015 Lausanne, Switzerland}

\author{Junda Wang}%
\affiliation{Laboratory of Wave Engineering, \'Ecole Polytechnique F\'ed\'erale de Lausanne (EPFL), CH-1015 Lausanne, Switzerland}

\author{Benjamin Apffel}
 \affiliation{Laboratory of Wave Engineering, \'Ecole Polytechnique F\'ed\'erale de Lausanne (EPFL), CH-1015 Lausanne, Switzerland}

\author{\\Aleksi Bossart}
 \affiliation{Laboratory of Wave Engineering, \'Ecole Polytechnique F\'ed\'erale de Lausanne (EPFL), CH-1015 Lausanne, Switzerland}
 
\author{Haoye Qin}
 \affiliation{Laboratory of Wave Engineering, \'Ecole Polytechnique F\'ed\'erale de Lausanne (EPFL), CH-1015 Lausanne, Switzerland}
 
\author{Oleg V. Yazyev }
\affiliation{Institute of Physics, \'Ecole Polytechnique F\'ed\'erale de Lausanne (EPFL), CH-1015 Lausanne, Switzerland}

\author{Romain Fleury}
\email{romain.fleury@epfl.ch}
\affiliation{Laboratory of Wave Engineering, \'Ecole Polytechnique F\'ed\'erale de Lausanne (EPFL), CH-1015 Lausanne, Switzerland}

%




\date{\today}

\begin{abstract}
\input{Abstract}
\end{abstract}

\maketitle


\section{\label{sec:Intro} Introduction}
\input{introduction}

\section{\label{sec:Scaling in scattering}  Scattering networks:  from microscopic to macroscopic scale}

\input{SectionII}

\section{\label{sec:RG I} RG on scattering networks with phase-link disorder}

\input{SectionIII}

\section{\label{sec:LL}Scaling analysis of the localization length}
\input{SectionIV}

\section{\label{sec:Exp}Experiments}
\input{SectionV}

\section{\label{sec:RG II}RG on scattering networks  with structural disorder}
\input{SectionVI}

\section{\label{sec:conclude} Conclusions and outlook}
\input{Conclusion}

\begin{acknowledgments}
We thank Andrei A. Fedorenko, Pierre Delplace, David Carpentier, Henning Schomerus, and Stefan Rotter for fruitful discussions. Z. Z., J. W., A. B., H. Q.  and R. F. acknowledge funding from the Swiss State Secretariat for Education, Research and Innovation (SERI) under contract number MB22.00028 and the Swiss National Science Foundation (SNSF) under the Eccellenza award 181232. Y. G.  and O. Y. acknowledge SNSF grant No. 204254.
\end{acknowledgments}


\appendix
\input{Appendix}

\bibliography{references}

\end{document}

%% file: Abstract.tex
Exploring and understanding topological phases in systems with strong distributed disorder requires developing fundamentally new approaches to replace traditional tools such as topological band theory. Here, we present a general real-space renormalization group (RG) approach for scattering models, which is capable of dealing with strong distributed disorder without relying on the renormalization of Hamiltonians or wave functions. Such scheme, based on a block-scattering transformation combined with a replica strategy, is applied for a comprehensive study of strongly disordered unitary scattering networks with localized bulk states, uncovering a connection between topological physics and critical behavior. Our RG scheme leads to topological flow diagrams that unveil how the microscopic competition between reflection and non-reciprocity leads to the large-scale emergence of macroscopic scattering attractors, corresponding to trivial and topological insulators. Our findings are confirmed by a scaling analysis of the localization length (LL) and critical exponents, and experimentally validated. The results not only shed light on the fundamental understanding of topological phase transitions and scaling properties in strongly disordered regimes, but also pave the way for practical applications in modern topological condensed-matter and photonics, where disorder may be seen as a useful design degree of freedom, and no longer as a hindrance.

%% file: Introduction.tex
\label{introduction}

Networks, discrete models comprising scatterers interconnected by links, can capture the key physics of phase transitions in complex physical systems involving the transport of waves or particles. Their use dates back to the early days of scaling theory \cite{stanley_scaling_1999,stauffer_scaling_1979,shapiro_renormalization-group_1982} and the quantum Hall effect (QHE) \cite{huckestein_scaling_1995,fertig_semiclassical_1988,cardy_network_2005,kramer_random_2005}. They include seminal models by Chalker and Coddington (CC) \cite{chalker_percolation_1988,ho_models_1996}, developed for describing the localization-delocalization transition in a quantum Hall system with a random potential. In a CC network, the equipotential lines of the disordered potential become unidirectional links over which particles travel, experiencing random phase delays. The probability to hop on another line is concentrated on the saddle points of the potential, represented in the network as nodes on which unitary scattering occurs. Such discretization to a network provides opportunities for quantitative predictions, especially via scaling analysis, as exemplified by the prediction of critical exponents in CC networks \cite{chalker_percolation_1988,kramer_random_2005} that agree with experimental data \cite{li_scaling_2005,slevin_critical_2009}. In the past several decades, fueled by the advent of topological physics, periodic unitary scattering networks have been widely studied from the standpoint of topology, demonstrating the possibility of obtaining helical boundary transport in quantum spin-Hall insulators \cite{obuse_two-dimensional_2007,obuse_spin-directed_2014}, or chiral edge states in Chern \cite{pasek_network_2014,delplace_phase_2017,zhang_superior_2021} and anomalous Floquet insulators \cite{pasek_network_2014,delplace_phase_2017,liang_optical_2013,titum_anomalous_2016,hu_measurement_2015,gao_probing_2016,liu_anomalous_2020,zhang_superior_2021,potter_quantum_2020}. 

Remarkably, recent works demonstrated that some of these chiral topological edge states can survive the addition of strong levels of distributed disorder in the network, in the form of arbitrary phase fluctuations, or even a randomization of its structure \cite{zhang_superior_2021,zhang_anomalous_2023,kim_floquet_2023,qin_anomalous_2023,chen_anomalous_2024}. This observation was validated by direct evaluation and experimental measurements of topological invariants in strongly amorphous cases \cite{zhang_anomalous_2023}. Yet, predicting whether a given periodic network will be topological or not when disorder is imparted remains a challenge \cite{shapiro_strongly_2019,graf_bulkedge_2018,prodan_disordered_2011}, and the reason why networks can retain a nontrivial topological nature in drastically aperiodic settings is still poorly understood. For example, some honeycomb topological networks supporting chiral edge states in this clean limit remain topological when adding disorder, whereas some others do not and trivially localize \cite{zhang_superior_2021,zhang_anomalous_2023,chen_anomalous_2024}. This behavior seems to be related to the networks being either in the anomalous or Chern phases in the clean periodic limit, although counter-examples can be found near topological phase transition boundaries. Such phenomena appear to indicate the existence of unexplored critical topological transitions in disordered networks, which by essence cannot be understood from standard approaches relying on topological band theories.

\begin{figure*}[htbp!]
\includegraphics[width=1\textwidth]{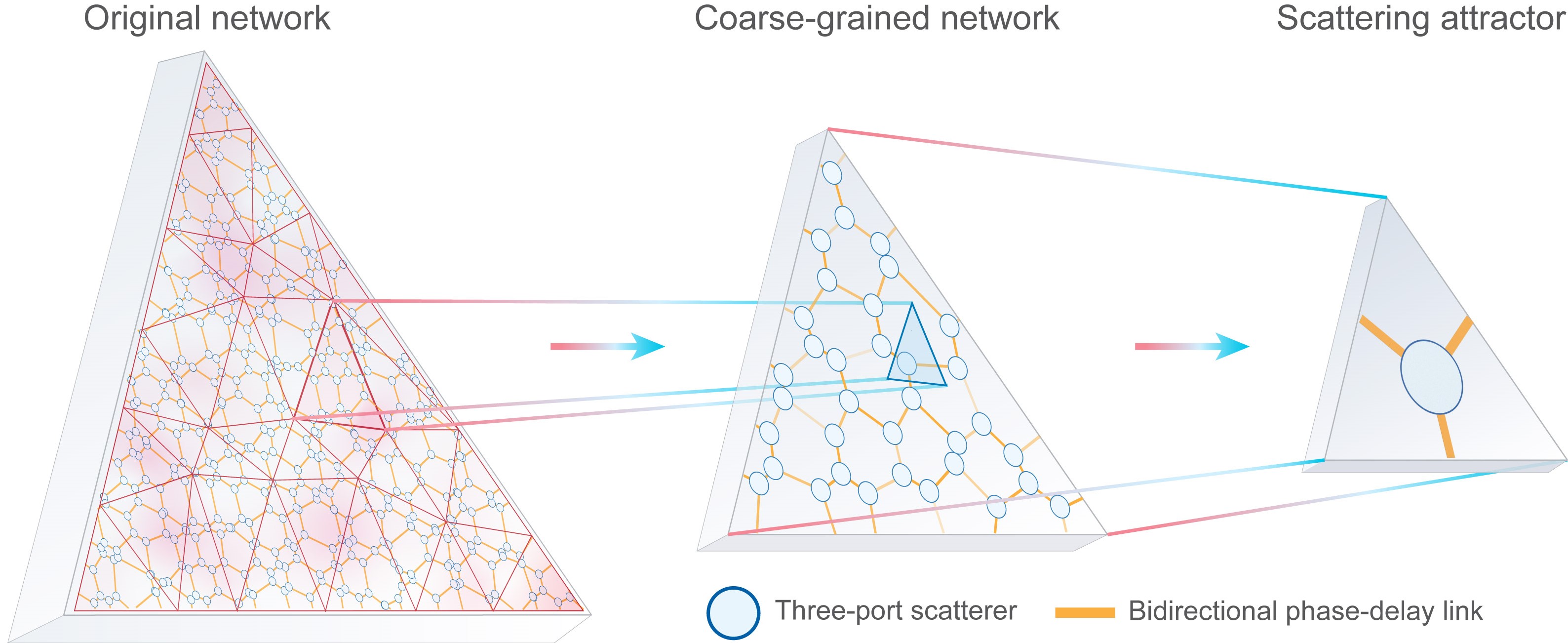}
\caption{\label{fig:Skech of RG- Intro} \textbf{Block scattering transformations for a renormalization group of unitary scattering networks.} Starting with an arbitrary scattering network composed of three-port unitary scatterers connected by reciprocal phase links (left), we perform block-scattering transformations to extract the key scattering properties of each triangular sub-blocks, which are replaced by a three-port unitary scattering matrix (center). This results in a coarse-grained network. By iteratively applying this transformation, we finally get to a single three-port scatterer described by a unitary 3 by 3 matrix $\boldsymbol{S}_F$, which we call the scattering attractor of the network (right). If the RG procedure is successful, the scattering attractor $\boldsymbol{S}_F$ summarizes whether the initial network is a trivial insulator (if $\boldsymbol{S}_F$ is a full reflection matrix)  or a topological insulator ($\boldsymbol{S}_F$ is a circulator).}
\end{figure*}




The renormalization group (RG) \cite{wilson_renormalization_1975,fisher_renormalization_1998,cardy_scaling_1996,efrati_real-space_2014}, which offers valuable insights on the connection between physical phenomena occurring at very different scales, may provide a way to probe the uncharted connection between scattering processes occurring at the microscopic scale, and macroscopic transport properties of large samples. The later are related to the scaling of insulating phases and their topology \cite{ludwig_integer_1994,arovas_real-space_1997,cain_real-space_2003,janssen_s-matrix_1998,schomerus_renormalization_2023} in periodic or aperiodic scenarios \cite{kosterlitz_ordering_1973}. RG is a conceptual frame: it catches large-scale behavior, predicting macroscopic physical observables while smearing out local fluctuations. Conceptual advances in the use of RG in condensed matter physics have led to important developments. For instance, recent works applied tensor-networks RG on state entanglement to describe symmetry-protected topological order \cite{gu_tensor-entanglement-filtering_2009,chen_local_2010}, perturbative RG on Hamiltonians to deduce the topological phase diagrams under local disorder \cite{morimoto_anderson_2015}, and momentum space-RG on Berry curvatures to identify topological transitions in periodic structures \cite{chen_scaling_2016}. However, RG approaches proposed in the context of topological physics have focused on either periodic systems, or systems with local disorder by renormalizing the Hamiltonians or the wave function. The case of topological transitions and scaling effects in systems with very strong non-local disorder, i.e. disorder of arbitrary strength distributed over their entire area, is still largely unexplored \cite{mong_quantum_2012,fulga_statistical_2014,prodan_non-commutative_2016,shapiro_strongly_2019}. Moreover, despite much interest in topological scattering networks and their potential applications in photonics \cite{hafezi_robust_2011,hafezi_imaging_2013,harris_quantum_2017,pai_parallel_2020,bogaerts_programmable_2020,afzal_realization_2020,dai_topologically_2022,wang_graph_2023,dai_non-hermitian_2024} and electromagnetic systems \cite{rohden_self-organized_2012,hul_experimental_2004,gao_probing_2016,xiang_intracity_2017}, a unifying RG scheme to understand these topological systems is still crucially lacking. RG on strongly disordered networks is expected to shed light on the competition between localization and topology in a broad range of scenarios, by revealing how microscopic scattering properties affect macroscopic topological transport.

\begin{figure}[htbp]
\includegraphics[width=0.48\textwidth]{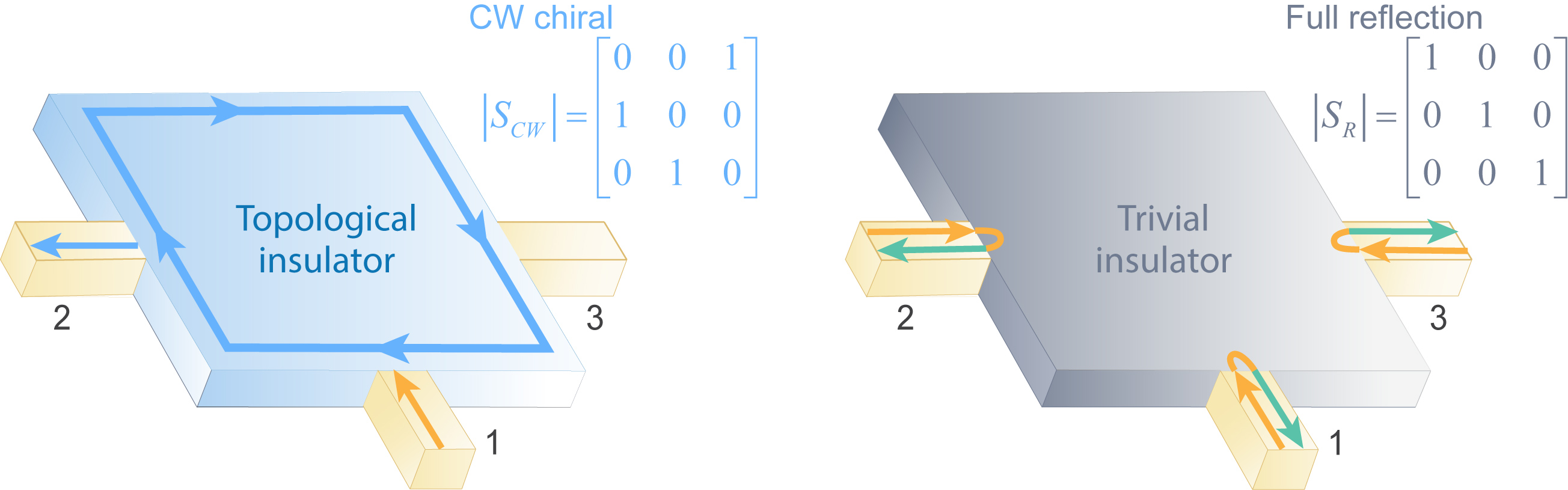}
\caption{\label{fig:S and Topo} \textbf{Trivial and topological systems and their scattering attractors.} We consider three matched probes (in yellow) placed at arbitrary positions on the boundary of a 2D insulator. This is the minimal number of probes allowing the detection of chiral edge transport. For a topological insulator (left, blue), we expect that as the system size increases, the scattering matrix at the probes converges to the one of a clockwise ($\boldsymbol{S}_{CW}$) or counter clockwise ($\boldsymbol{S}_{CCW}$) unitary circulator. On the other hand, the probe scattering matrix for an ideal trivial insulator (right, gray) would converge to a full reflection matrix $\boldsymbol{S}_R$. Therefore, we expect $\boldsymbol{S}_{CW}$ ($\boldsymbol{S}_{CCW}$) and $\boldsymbol{S}_R$ to represent possible scattering attractors in any valid RG scheme.}
\end{figure}

In this article, we propose a real-space renormalization group on unitary scattering networks, which unveils the intricate physical mechanisms behind the persistence of topological edge states in systems with strong distributed disorder. Instead of playing with Hamiltonians or wave functions, we focus on network models and propose block-scattering transformations that preserve the key scattering properties of each block during scaling, namely flux conservation, reflection level, and scattering chirality. The block-scattering transformation (Fig. \ref{fig:Skech of RG- Intro}) is composed of three steps: partitioning the original network triangularly, replacing each block subnetwork by a simpler three-port scatterer, and interconnecting together these three-port scatterers into a new coarse-grained network, on which the procedure can be repeated.  The goal of the RG scheme is therefore to leverage iterative block-scattering transformations until one converges to a three-port scattering attractor that captures the essential information about the macroscopic scattering of the network, namely whether chiral transport occurs on the edges or if incident waves coupled to the edge are just reflected. Intuitively, we expect three possible scattering attractors (Fig. \ref{fig:S and Topo}), which should correspond to stable fixed points of the RG flow. Chiral topological systems would be attracted to unitary scattering matrices that describe clockwise or counter clockwise perfect circulation. The clockwise circulator $\boldsymbol{S}_{CW}$ (matrix shown in the figure inset), and its transpose $\boldsymbol{S}_{CCW}$ are the only two possibilities compatible with edge transport (we ignore transmission phases for now). On the other hand, systems that trivially localize would be attracted to the identity matrix  $\boldsymbol{S}_R$ (we ignore reflection phases for now). This later case corresponds to full reflection as the input waves excite localized modes. We apply this RG scheme on two examples of fully disordered networks, either with a honeycomb structure subject to arbitrary phase fluctuations on the hexagonal links, or a fully random structure with arbitrary planar connectivity. We obtain RG topological phase diagrams that elucidate the intricate competition between microscopic reflection and chirality. We unveil the critical phenomena occurring at the transition between trivial and topological disordered networks, by exploring the evolution of RG flows upon scaling and studying the critical probability distributions of microscopic scattering matrices. This block-scattering RG approach leads to a better understanding of topological phase transitions and scaling properties in scattering networks models with strong disorder, broadening the scope of renormalization group approaches to topological unitary systems.

To validate the accuracy of our scattering RG scheme, we confront it to the results of a direct scaling analysis of the localization length (LL), which provides critical boundaries together with their critical exponents. The critical boundaries obtained from LL scaling analysis match well the phase transition boundaries independently obtained from RG, confirming the accuracy of our block scattering matrix transformations. The resulting critical exponents $\nu$ exhibit two values on the boundary (around 2.43 and 3.33), corresponding to different unstable fixed points/lines on the RG flows, originating from disparate types of topological phase transitions. Furthermore, we perform experiments allowing us to directly measure RG flows when scaling microwave scattering networks with phase-link disorder. The measured RG flow is in agreement with the one predicted by theory. In the case of structurally disordered networks, our RG analysis evidences a slightly contracted topological phase region when compared to phase-link disordered honeycomb networks, fully elucidating the puzzling examples of topological and trivial networks that initially motivated this study.

The paper is organized as follows: In Sec. II, we introduce different scattering network configurations and distributed disorder types, and show examples introducing the problem of predicting the emergence of chiral edge states upon scaling. Sec. III, IV, V focus on networks with strong phase-link disorder, describing the scattering RG scheme and results (Sec. III), its validation by scaling analysis of the localization length (Sec. IV), and an experimental validation of RG flows (Sec. V). In Sec. VI, we extend the RG analysis to structural network disorder. In Sec. VII, we conclude our results and discuss their implications. Appendices give details on the block-scattering transformation and how the RG scheme is numerically implemented, as well as details on clean-limit topological phase diagrams, critical behavior and localization length calculations.


%% file: SectionII.tex
\label{SectionII}
\begin{figure*}[htbp!]
\includegraphics[width=0.8\textwidth]{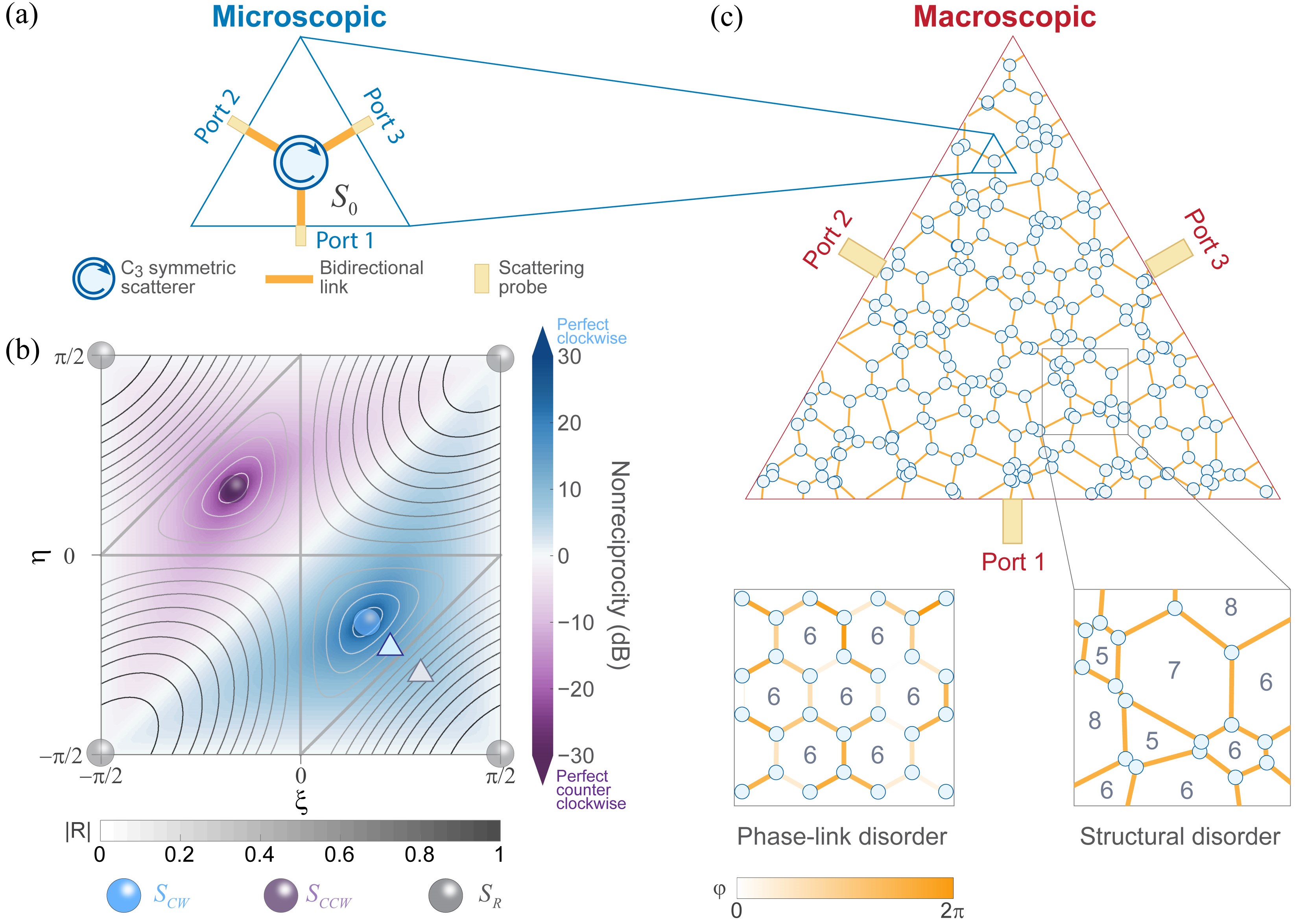}
\caption{\label{fig:scale_scattering}\textbf{Scattering networks, from the microscopic to the macroscopic scale. }(a) Unitary microscopic scatterers with three ports and three-fold rotational symmetry are used as building blocks to form macroscopic networks. The scattering detected at three scattering probes is described by the unitary matrix $S_0$. (b) Reflection (contour lines) and nonreciprocity (color map) of $S_0$ in its parameter space, defined by the two angle parameters $(\xi, \eta) \in [-\pi/2, \pi/2) \times [-\pi/2, \pi/2)$. See the main text for a definition of these quantities. Three special scattering matrices $\boldsymbol{S}_R$, $\boldsymbol{S}_{CW}$, $\boldsymbol{S}_{CCW}$, corresponding respectively to a fully reflective scatterer, a clockwise perfect circulator, and a counter-clockwise perfect circulator, are located on this plane. (c) Macroscopic scattering network made of microscopic three-port scatterers $S_0$ interconnected by bidirectional phase-delay links. Different types of distributed disorder can be present. A first class of disorder consists in taking a periodic network, e.g. a honeycomb arrangement, and add random phase-delay fluctuations on each links (bottom left panel). A different kind of disorder can take the form of deformations of the network structure, completely breaking the hexagonal structure (bottom right panel, numbers count the number of sides forming each loop). The macroscopic scattering properties of networks are defined at three external probes, and dictated by microscopic properties, in particular the value of $S_0$.}
\end{figure*}

At the microscopic scale, the networks we consider are formed of interconnected unitary scatterers with three ports (Fig. \ref{fig:scale_scattering}(a)). Such structure is maintained through renormalization iterations until converging to the attractor, which is again a three-port unitary scatterer. This choice to work with three-port systems is motivated by two reasons. First, at the microscopic scale, scatterers with more than two ports are needed to construct complex networks. Second, at the macroscopic scale of the attractor, it wouldn't be possible to detect chiral edge transport with only two ports, and three-port appears here as a minimal number to do so \cite{guo_reciprocity_2022}. One could, of course, build a theory based on four-port unitary scattering, but this would only complicate the associated mathematics without bringing any new advantage. Yet, one could object the following: what if the initial network contains scatterers with arbitrary numbers of ports? Well, in this case, we can always make a first coarse-grained iteration, and this operation would reduce all subsequent iterations back to the three-port case. We will also assume that the individual scatterers that compose the initial networks obey three-fold rotational ($C_3$) symmetry. However, we will assume that $C_3$ symmetry is generally broken by the entire network, as the connections between the scatterers do not fulfill this symmetry. This means that after the first RG iteration, the scattering matrices in the coarse grain picture are no longer $C_3$ symmetric. Nevertheless, for representing the evolution of these matrices during RG iterations, we will perform ensemble averaging, which will restore $C_3$ symmetry for sufficiently large statistical ensembles. For all these reasons, three-port scatterers described by a $C_3$ and  unitary ($U(3)$) scattering matrix $\boldsymbol{S}_0$ play a crucial role in our scheme. Such complex-valued matrices take the form
\begin{equation}
\boldsymbol{S}_0=
\begin{bmatrix}
R & T_{CCW} & T_{CW}\\
T_{CW} & R & T_{CCW}\\
T_{CCW} & T_{CW} & R
\end{bmatrix},
\label{S0}
\end{equation}
where $T_{CW}$ and $T_{CCW}$ represent chiral clockwise (CW) and counter clockwise (CCW) transmissions, respectively, while $R$ is the reflection. Such parametrization is necessary to enforce $C_3$ symmetry, but not sufficient to guarantee unitarity, which implies additional orthogonality constraints on these coefficients. Forgetting about a global phase, these constraints can be elegantly satisfied if one retains a parametrization of these coefficients in terms of only two angles, $\xi$ and $\eta$, both of which are defined in the interval $[-\pi/2, \pi/2)$ \cite{zhang_superior_2021,zhang_topological_nodate}. With $\xi$ and $\eta$ parameters, $R, T_{CW}, \text{and }T_{CCW}$ are expressed by
\begin{align} 
R(\xi,\eta) &=-1+\frac{2}{3} \cos{\xi} e^{i\xi}+\frac{2}{3} \cos{\eta} e^{i\eta} \\
T_{CW}(\xi,\eta) &=\frac{2}{3} \left[ e^{i \frac{2}{3} \pi}\cos{\xi} e^{i\xi}+e^{-i \frac{2}{3} \pi} \cos{\eta} e^{i\eta} \right] \\
T_{CCW}(\xi,\eta) &=\frac{2}{3} \left[ e^{-i \frac{2}{3} \pi}\cos{\xi} e^{i\xi}+e^{i \frac{2}{3} \pi} \cos{\eta} e^{i\eta} \right].
\label{Parametrezation_Details}
\end{align}

\begin{figure*}[htbp!]
\includegraphics[width=0.9\textwidth]{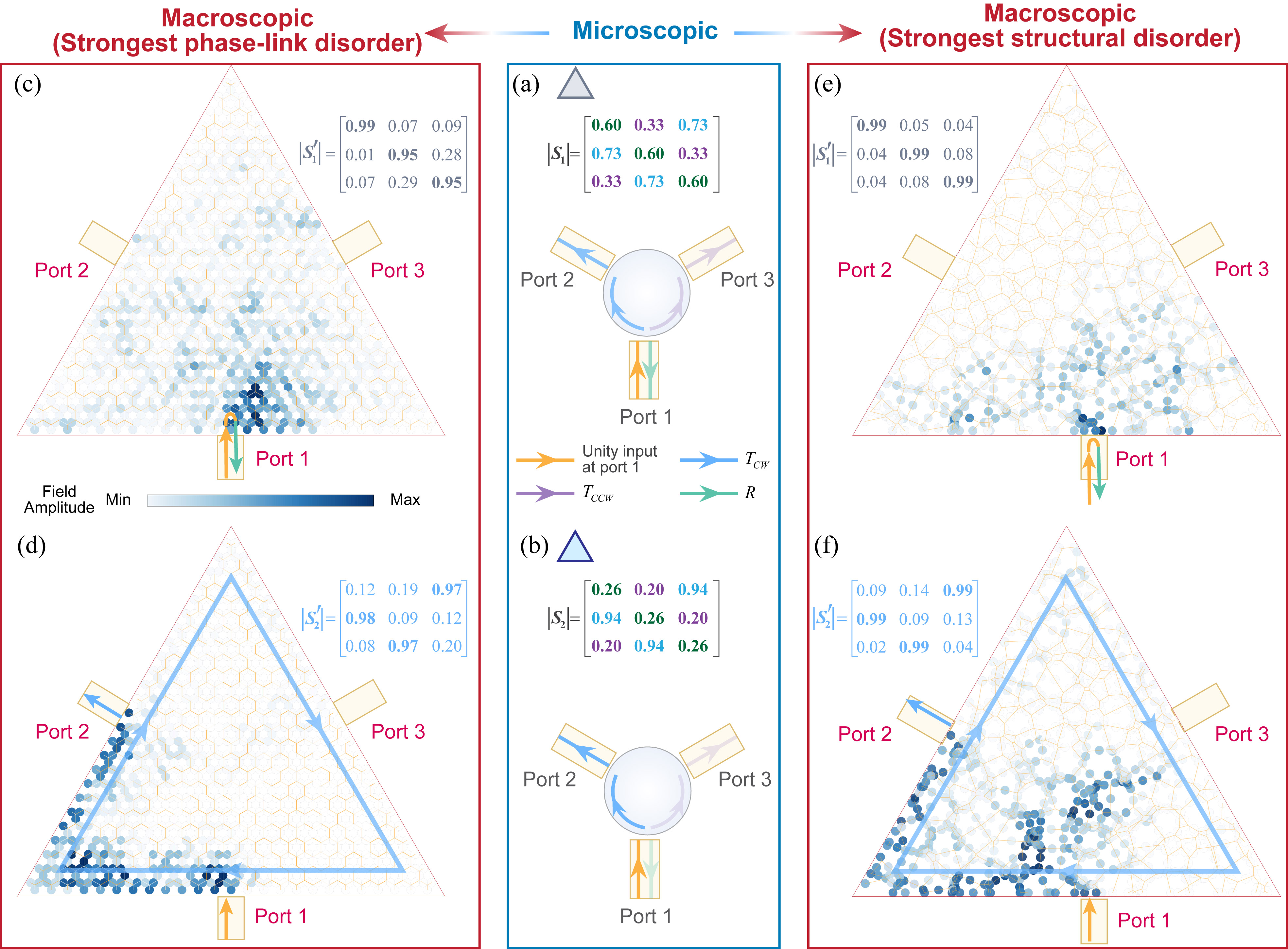}
\caption{\label{fig:finiteexamples} \textbf{Examples leading to trivial and topological macroscopic properties}, under phase-link disorder and structural disorder. (a) and (b) We choose two microscopic scattering matrices $\boldsymbol{S}_1$ (a) and $\boldsymbol{S}_2$ (b), located at $\xi= -\eta= 2.8 \pi/12$ (gray triangle) and $\xi= -\eta= 3.8 \pi/12$ (blue triangle) on the parameter plane, as marked in Fig. \ref{fig:scale_scattering}(b), respectively. Scattering flows are shown by coloured arrows, illustrating $T_{CW}$ (blue), $T_{CCW}$ (purple), and $R$ (green), when a unity wave is incident from port 1 (yellow arrow). $\boldsymbol{S}_1$ and $\boldsymbol{S}_2$  both exhibit clockwise chirality with some reflection, and are relatively close to each other on the parameter plane. These matrices are then used to form networks with maximal phase-link disorder (c-d) or structural disorder (e-f). We computed field maps when inputting a signal at port 1, and calculated the associated macroscopic scattering matrices. $\boldsymbol{S}_1$ always leads to a trivial insulator with full external reflection, whereas $\boldsymbol{S}_2$ yields topological edge states in both cases.}
\end{figure*}

The possible scattering attractors $\boldsymbol{S}_R$ and $\boldsymbol{S}_{CW}$ ($\boldsymbol{S}_{CCW}$) are special cases that belong to this family of matrices. For example, as shown in Fig. \ref{fig:scale_scattering}(b), the fully localized matrix $\boldsymbol{S}_R=\boldsymbol{I}$ is obtained when $\xi$ and $\eta$ are equal to $\pm \pi/2$, whereas the chiral transport matrix $\boldsymbol{S}_{CW}$ ($\boldsymbol{S}_{CCW}$) is achieved under $\xi= -\eta= \pi/6$ ($-\pi/6$). By varying $\xi$ and $\eta$, we can generate all $C_3$ symmetric $U(3)$ scattering matrices $\boldsymbol{S}_0(\xi,\eta)$ with variable reflection $R(\xi,\eta)$ and scattering chirality, which is quantified by the nonreciprocity level $NR(\xi,\eta)= T_{CW}/T_{CCW}$. This family of matrices is represented in Fig. \ref{fig:scale_scattering}(b), where the coloured density map represents $NR(\xi,\eta)$ in decibels, and the grey contours are for $R(\xi,\eta)$. The scattering attractors are located at global extrema of $R(\xi,\eta)$ and $NR(\xi,\eta)$ on this parametric plane. 

Macroscopic planar networks are then built by connecting such $\boldsymbol{S}_0$ scatterers using bidirectional links. One can think of these links as lossless monomode waveguides characterized by the phase delay $\varphi$ that they impart to waves traveling one time along their length, in any direction. An example of a structurally disordered macroscopic network is shown in Fig. \ref{fig:scale_scattering}(c). Among other possible macroscopic networks, a special configuration is the one of periodic honeycomb networks, for which identical phase-delay links are used. We will refer to honeycomb networks as pristine or clean-limit scattering networks, because it is an archetypal case in which no disorder is present. Such periodic clean limit has the advantage of exhibiting well-understood topological phases, extensively described in prior arts \cite{zhang_superior_2021}. In particular, two regions centered on $\boldsymbol{S}_{CW}$  and $\boldsymbol{S}_{CCW}$ correspond to anomalous Floquet topological phases with clockwise and counter-clockwise chiral edge states, respectively (see Appendix \ref{ap:cleanlimit} for details on the topological phase diagram of pristine honeycomb networks).

In this paper, we focus instead on network models subject to disorder that is uniformly distributed over their entire area. We consider two types of disorder. First, phase-link disorder can be imparted on the bi-directional links without changing the honeycomb structure of the network, and is represented by a probability distribution of phase-delay values $P(\varphi)$. This is illustrated in the bottom left inset of Fig. \ref{fig:scale_scattering}(c). Second, one can imagine changing the structure of the network, creating in it loops that are no longer regular hexagons, but arbitrary irregular polygons (bottom right inset). Such structural disorder can be continuously added to the honeycomb clean limit, following known Voronoi tessellation techniques \cite{marsal_topological_2020,zhang_anomalous_2023,grushin_amorphous_2023,cassella_exact_2023}. An important point is that this addition of structural disorder is controlled by a single parameter, the amorphous factor $\alpha$. Both types of disorder can be made maximally strong: phase disorder is purely random when $P(\varphi)$ is a uniform distribution in $[0, 2\pi)$, while the maximal amorphism attainable on a Euclid plane is reached when $\alpha\geq 6$ \cite{zhang_anomalous_2023}.  In this work, we will study both types of disorder, assuming always maximally strong levels. The macroscopic scattering networks are then probed via three external scattering probes (Fig. \ref{fig:scale_scattering}(c)), defining a macroscopic unitary scattering matrix $\boldsymbol{S}_0^{'}$. The first goal of RG is to establish a link between the microscopic scattering $\boldsymbol{S}_0$  and the macroscopic one $\boldsymbol{S}_0^{'}$, and unveil the role of disorder in this mapping. The second goal of RG is identifying the limit of $\boldsymbol{S}_0^{'}$ when the macroscopic system gets infinitely large (we call this limit the thermodynamic limit), to unveil the relation between scattering attractors and topological phases. This is made possible by the fact that $\boldsymbol{S}_0^{'}$ can contain the signature of topological edge transport, even under strong disorder \cite{fulga_scattering_2011,fulga_scattering_2012,franca_non-hermitian_2022}.

The connection between $\boldsymbol{S}_0$ and $\boldsymbol{S}_0^{'}$ is by no mean trivial. Some numerically tractable examples with relatively large sizes are shown in Fig. \ref{fig:finiteexamples}. We start with panels (a-b) in the center of the figure, by choosing two microscopic building blocks, whose scattering matrices $\boldsymbol{S}_1$ and $\boldsymbol{S}_2$ are given in the figure. We then build macroscopic scattering networks with either phase-link disorder (panels (c-d) on the left) and structural disorder (panels (e-f) on the right). Field maps are computed numerically assuming input from the bottom port. Note that $\boldsymbol{S}_1$ and $\boldsymbol{S}_2$, which are marked by grey and blue triangles in Fig. \ref{fig:scale_scattering}(b), are both of the same chirality and differ only slightly in their level of reflection and nonreciprocity. Despite of this, we observe a completely opposite macroscopic scattering behavior in the networks originating from $\boldsymbol{S}_1$ and $\boldsymbol{S}_2$. In the case of $\boldsymbol{S}_1$ (Fig. \ref{fig:finiteexamples}(c, e)), the input wave seems to localize around the input port and end up being reflected. Conversely, with $\boldsymbol{S}_2$ a clear edge transport channel appears and connects the input port to the next port on the left. If we could numerically access the thermodynamic limit, we would expect a convergence of the macroscopic scattering matrices $\boldsymbol{S}_1^{'}$  and $\boldsymbol{S}_2^{'}$ to $\boldsymbol{S}_R$ and $\boldsymbol{S}_{CW}$ (the values of $\boldsymbol{S}_1^{'}$ and $\boldsymbol{S}_2^{'}$ for the finite systems considered here are shown within the figure). We also found even more surprising examples of systems behaving oppositely for the two types of disorder, when starting from scattering matrices with reflectance slightly lower than $1/3$ (not shown for brevity). Explaining the emergence or non-emergence of unidirectional edge states in networks with distributed disorder requires a unified scheme capable of accessing the macroscopic properties of arbitrarily large systems with no additional computational cost. In the next section, we describe a RG scheme that sheds light on these examples and unveils a microscopic competition between reflection and chirality, that translates into a macroscopic competition between disorder and topology, explaining the emergence of critical behaviors associated to topological transitions in the thermodynamic limit.

%% file: SectionIII.tex
\label{SectionIII}
\subsection{\label{sec:Frame}Iterative block-scattering transformations}
\begin{figure*}[htbp]
\includegraphics[width=0.9\textwidth]{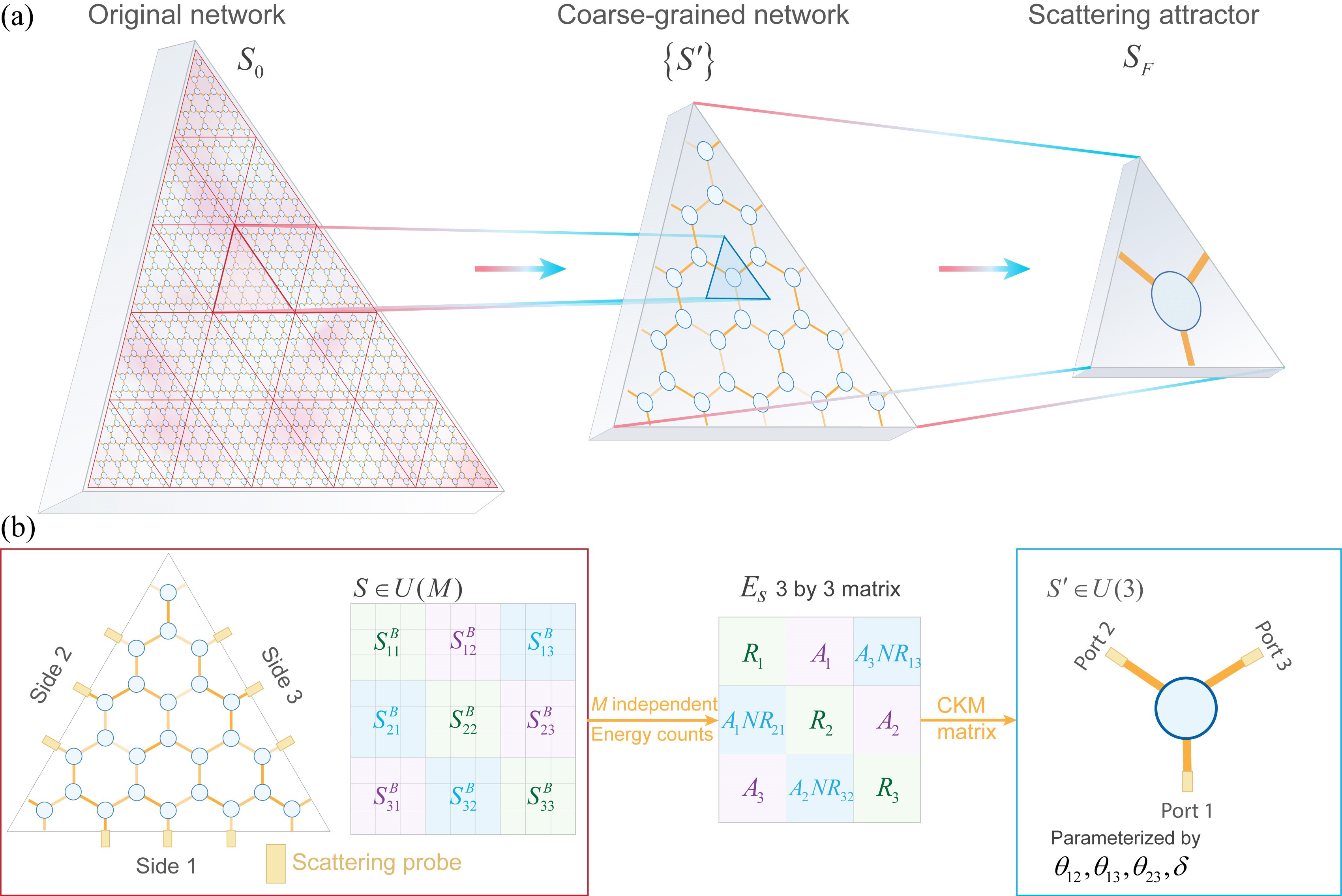}
\caption{\label{fig:RGI}\textbf{Renormalization group of a unitary scattering network with phase-link disorder. }(a) Schematic of the block-scattering transformation. (b) Procedure for the replacement of a triangular block scattering network (described by a large unitary scattering matrix $\boldsymbol{S}\in U(M)$) by a single three-port scatterer (described by a 3×3 scattering matrix $\boldsymbol{S}^{\prime}\in U(3)$). The reduction of matrix size is performed by partitioning $\boldsymbol{S}$ into nine 3×3 blocks according to the three sides, then summarizing some key dimensionless quantities, namely \textit{nonreciprocity and reflection} into an energy matrix $\boldsymbol{E}_S$ (center panel). $A_i (i=1, 2, 3)$ are variables to be determined in order to restore unitarity. This is done using CKM matrix parameterization, and we recover the corresponding unitary matrix $\boldsymbol{S}^{\prime}\in U(3)$ (rightmost panel).}
\end{figure*}
The type of phase-disordered networks we focus on in this section is shown in Fig. \ref{fig:RGI}. They are composed of identical scattering matrices $\boldsymbol{S}_0(\xi,\eta)$ connected in a honeycomb structure with phase-link disorder. The phase $\varphi$ of each link is drawn from a uniform distribution in the range $[\varphi_0-\Delta \varphi/2, \varphi_0+\Delta \varphi/2]$ (Fig. \ref{fig:RGI}). We also focus on the case of the strongest possible disorder level $\Delta \varphi= 2\pi$, although any other range is in principle accessible.  As discussed in the introduction, the RG scheme contains three steps. First, we subdivide the original network into triangular blocks by applying a standard Delaunay  triangulation, whose generators are on a triangular lattice. Second, each triangular block is replaced by a single three-port scatterer, whose scattering matrix $\boldsymbol{S}^{\prime}\in U(3)$ must be deduced from the scattering properties of the block. In the final step, by considering the dual graph of the triangulation graph, we can form a coarse-grained network by arranging the matrices $\boldsymbol{S}^{\prime}$ on the nodes, and connect them with new phase links randomly drawn from the same uniform distribution as the original ones. The RG transformation is then iterated, replacing the original network by the coarse-grained one. Each iteration reduces the size of the network by a scaling factor equal to the number of scatterers $P$ in the triangular blocks ($P$= 25 in panel a). After $n$ iterative transformations, a 3-port scatterer therefore represents $25^n$ scatterers in the original network. At the end, we get a single three-port attractor $\boldsymbol{S}_{F}\in U(3)$.

The backbone of the RG scheme is therefore the block-scattering transformation, namely the protocol of replacing the large unitary scattering matrix of a triangular block by a small one. Two pivotal questions arise when trying to define such a protocol: ``What scattering matrix should be associated to a triangular block ?" and, ``What kind of physical information should be preserved when this block scattering matrix is compressed into a small 3×3 matrix?". To address the first question, we note that a triangular block may have a large number of ports on its edges, which at the end should be concatenated into a single one. A simple proposition to achieve this concatenation would be to close all edge ports with full-reflections, leaving only three of them, one on each edge. However, such a solution would not consider the fact that the connections between adjacent triangular blocks are actually distributed over many ports along the edge, allowing to couple together various modes of such triangles. To take this into account, we do not select only one, but $M_i$ ports along the boundary of the $i$th side of the triangle, as shown on the left of Fig. \ref{fig:RGI}(b). On each side, these open ports are chosen to be adjacent, and we avoid choosing the ones around the corners. All the other ports are closed with a fully reflective boundary condition. This allows defining a unitary scattering matrix $\boldsymbol{S}\in U(M)$, with $M=\sum_{i=1}^{3} M_i$, which summarizes the transport and reflection at the scale of a block. Therefore, the problem of replacing the block scattering network by a three-port scatterer is equivalent to reducing the large unitary matrix $\boldsymbol{S}\in U(M)$ to a much smaller one, $\boldsymbol{S}^{\prime}\in U(3)$.

This takes us to our second pivotal question. To preserve the important physics while smearing out microscopic details, such reduction of scale must carefully maintain the scattering properties that matter in the trivial or topological localization processes occurring during scaling. Intuitively, the level of \textit{nonreciprocity and reflection} of a triangular block are important. We evidence these two properties in $\boldsymbol{S}$ by partitioning it into nine 3×3 blocks of sizes $M_i$, according to the side of the triangle, 
\begin{equation}
\boldsymbol{S}=
\begin{bmatrix}
\boldsymbol{S}_{11}^{B} & \boldsymbol{S}_{12}^{B} & \boldsymbol{S}_{13}^{B}\\
\boldsymbol{S}_{21}^{B} & \boldsymbol{S}_{22}^{B} & \boldsymbol{S}_{23}^{B}\\
\boldsymbol{S}_{31}^{B} & \boldsymbol{S}_{32}^{B} & \boldsymbol{S}_{33}^{B}
\end{bmatrix}.
\end{equation}
Next, we note that the reduction of $\boldsymbol{S}\in U(M)$ to $\boldsymbol{S}^{\prime}\in U(3)$ should follow a few principles. The first one is the fact that the quantities that we are trying to keep during the reduction should not depend on the size of the block network or the number of probes we choose. The second principle is that we should reflect accurately the way with which energy incident on one side of the triangular block is reflected and transmitted to the other two sides. One may start, following Landauer-Büttiker formalism \cite{buttiker_four-terminal_1986,rotter_light_2017}, by expressing the overall energy transport from side $i$ to side $j$ as
\begin{equation}
T_{ji}= \mathrm{Tr} (\boldsymbol{S}_{ij}^{B}(\boldsymbol{S}_{ij}^{B})^{\dag}),
\end{equation}
however this quantity depends on block size and number of probes. Instead, we can consider the overall nonreciprocity of the energy transport between side $i$ and $j$, represented by 
\begin{equation}
NR_{ji} = T_{ji}/T_{ij}.
\end{equation}
The overall reflection $R_{i}$ for the side $i$ of a triangular block may be represented by the quadratic mean of all the reflection of  the probes on this side, expressed as 
\begin{equation}
R_{i}= \sqrt{\sum_{p=1}^{M_i} |\boldsymbol{S}_{ii}^{B}(p,p)|^2/M_i}.
\end{equation}
After extracting the nonreciprocity and reflection of the block, $NR_{ji}$ and $R_{i}$, we can summarize this information about $\boldsymbol{S}$ into a 3×3 matrix $\boldsymbol{E}_S$ (Fig. \ref{fig:RGI} (b), center), defined as
\begin{equation}
\boldsymbol{E}_S=
\begin{bmatrix}
R_{1} & A_{1} & A_{3}\cdot NR_{13}\\
A_{1}\cdot {NR}_{21} & R_{2}& A_{2} \\
A_{3} & A_{2}\cdot NR_{32} & R_{3}
\end{bmatrix},
\label{eq:ES}
\end{equation}
where $NR_{ji}$ and $R_{i}$ serve as the non-diagonal and diagonal terms respectively. Here, $A_i (i=1, 2, 3)$ are variables to be determined so that we can recover a genuine unitary scattering matrix from $\boldsymbol{E}_S$. Note that $\boldsymbol{E}_S$, by itself, is not unitary and contains only amplitude information. The recovery of a unitary matrix from the matrix $\boldsymbol{E}_S$ is not a trivial task. Indeed, the general parameterization of $U(3)$ matrices implies that we have to recover in general three angle parameters ($\frac{1}{2}n(n-1)|_{n=3}$) and six phase parameters ($\frac{1}{2}n(n+1)|_{n=3}$) \cite{murnaghan_orthogonal_1958,reck_experimental_1994,dita_parametrisation_1994,dita_factorization_2001}). Fortunately, the problem of recovering a unitary scattering matrix from amplitude measurements is a known problem in high energy physics, solved by the Cabibbo–Kobayashi–Maskawa (CKM) matrix parameterization \cite{chau_comments_1984,dita_separation_2006}. This method implies that the recovery of a unitary matrix from an energy-related matrix can focus on finding four parameters (three angle parameters $\theta_{12},\theta_{13},\theta_{23}$ and one phase parameter $\delta$), as exemplified by the successful recovery of the quark-mixing matrix from experimental measurements. Appendix \ref{ap:replacement} details how to adapt CKM parametrization to the recovery of one possible candidate of a 3×3 unitary matrix $\boldsymbol{S}^{\prime}\in U(3)$, by first transforming $\boldsymbol{E}_S$ to a double stochastic matrix. We stress that many different choices for $\boldsymbol{S}^{\prime}\in U(3)$ are possible. However, they all differ by different phases, whose particular choice does not matter for RG due to the fact that random phase disorder is anyways present in the network at each iteration.

\begin{figure*}[htbp!]
\includegraphics[width=0.9\textwidth]{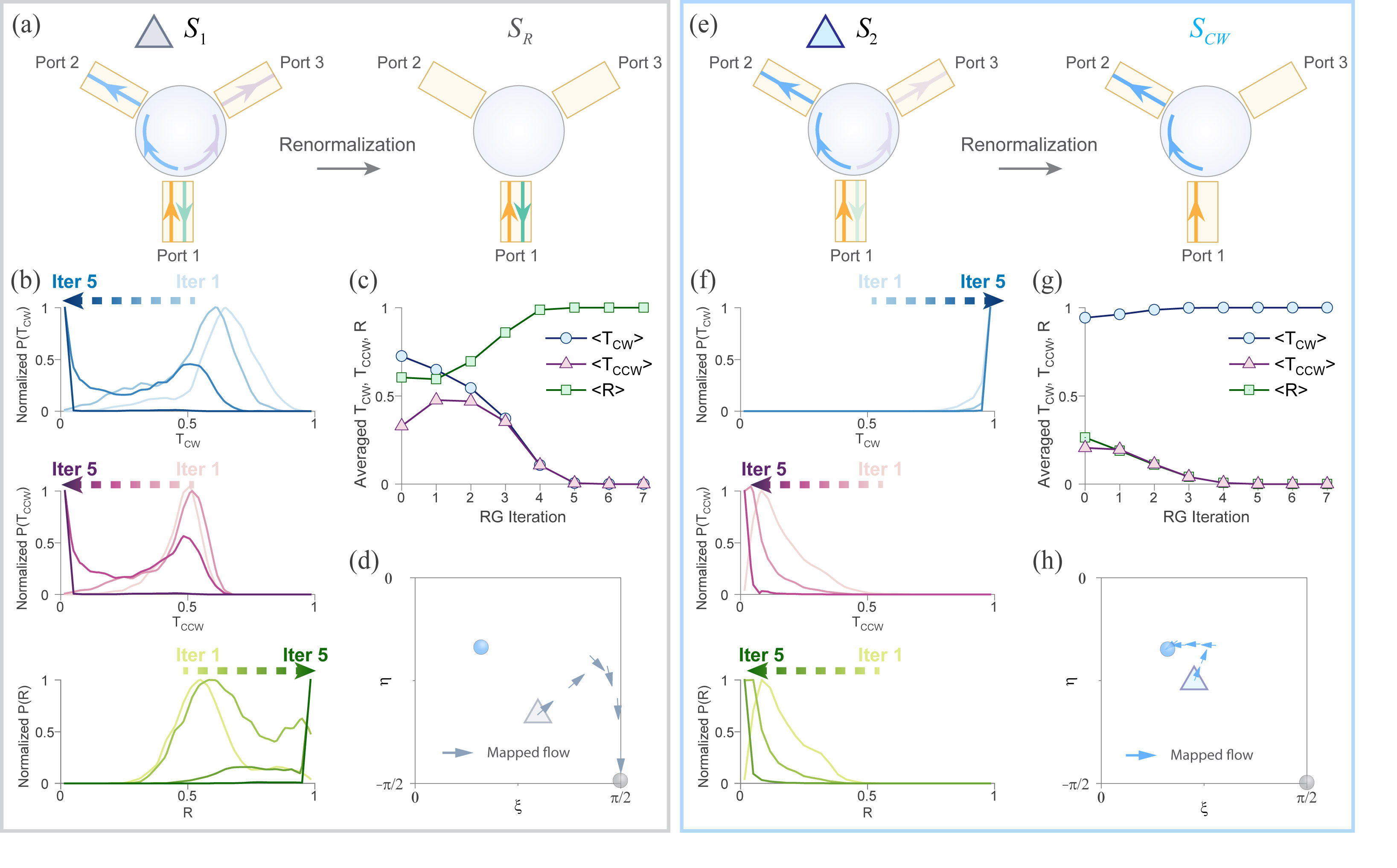}
\caption{\label{fig:two examples}\textbf{Two examples of RG on scattering networks with random phase-link disorder.} (a) and (e) Two opposite scattering attractors are reached for  $\boldsymbol{S}_1$ (a) and  $\boldsymbol{S}_2$ (e).  During RG, macroscopic disordered networks built from $\boldsymbol{S}_1$ and  $\boldsymbol{S}_2$ converge to the trivial attractor $\boldsymbol{S}_R$, and the clockwise chiral one $\boldsymbol{S}_{CW}$, respectively. (b) and (f) Evolution of the distributions $P(T_{CW})$, $P(T_{CCW})$, and $P(R)$ upon RG, for the case of $\boldsymbol{S}_1$ (b) and the case of $\boldsymbol{S}_2$ (f). These distributions represent how the macroscopic scattering properties continuously evolve when scaling up the networks. In panel (c) and (g), we summarize the distinct behaviors of $\boldsymbol{S}_1$ and $\boldsymbol{S}_2$ networks by plotting the statistic averages of these distributions upon scaling, $\langle T_{CW} \rangle$, $\langle T_{CCW} \rangle$, and $\langle R \rangle$. (d) and (h) RG flows. The averaged scattering properties of networks at each iteration are mapped to a point on the parameter space introduced with Fig. \ref{fig:RGI}(b). The trajectory of the point forms the RG flow, showing the evolution of the macroscopic scattering properties as the network is repeatedly scaled up during RG.}
\end{figure*}
\subsection{\label{sec:Replica}Replica scheme}
Having explained how the block scattering transformation is implemented, we are in principle ready to look at results from applying RG to large networks. Before doing so, however, one needs to think about two important practical aspects: (i) how to best describe the state of the network at a given iteration; and (ii), how to ensure that the iterative RG algorithm is computationally efficient. The first point is addressed by remembering that we want to track the convergence to a potential scattering attractor. Therefore, what matters is to monitor the evolution of the set of  $U(3)$ scattering matrices $\bigl\{\boldsymbol{S}_{n}^{\prime}\bigr\}$ during RG. This can be performed by averaging them over the network. This average is trivial to take for the initial network, which is composed of identical $C_3$ and $U(3)$ matrices $\boldsymbol{S}_0$. At any other iteration number $n$, the $\boldsymbol{S}^{\prime}_{n}$ in the network are in principle all different and no longer obey $C_3$ symmetry. However, we remark that averaging over a sufficiently large set would restore $C_3$ symmetry, since the choice of the port labels (1 , 2 and 3) for each $\boldsymbol{S}^{\prime}_{n}$ is arbitrary (as long as they are clockwise), eventually allowing for the data to be tripled, making the average invariant with respect to 120 degrees rotations. Thus, a first way to represent the state of the network at a given iteration is a point representing this average in the diagram of Fig. \ref{fig:scale_scattering}(b). This is equivalent to say that the statistic average of the scattering properties $\bigl\{\boldsymbol{S}_{n}^{\prime}\bigr\}$ in the $n_{th}$ coarse-grained network takes the form of a $C_3$ symmetric unitary matrix, namely
\begin{equation}
\langle \boldsymbol{S}_n^{\prime} \rangle \equiv
\begin{bmatrix}
\langle R_n \rangle & \langle T_{CCW,n} \rangle & \langle T_{CW,n} \rangle\\
\langle T_{CW,n} \rangle & \langle R_n \rangle & \langle T_{CCW,n} \rangle\\
\langle T_{CCW,n} \rangle & \langle T_{CW,n} \rangle & \langle R_n \rangle
\end{bmatrix}.
\label{eq:<S'n>}
\end{equation}
During the iterative RG procedure, the block-scattering transformation of the system from one scattering state $\langle \boldsymbol{S}_{n-1}^{\prime} \rangle$ to the next $\langle \boldsymbol{S}_n^{\prime} \rangle$ can thus be represented by the motion of this point, discretely jumping on the $(\xi,\eta)$ plane of Fig. \ref{fig:scale_scattering}(b). Starting from the point representing the original scatterer $\boldsymbol{S}_0$ on the $(\xi,\eta)$ plane, we can watch the trajectory formed by the successive locations of $\langle \boldsymbol{S}_n^{\prime} \rangle$, which defines a renormalization-group flow \cite{wilson_renormalization_1975}. Such flow can either lead the system towards a stable fixed point (a scattering attractor), the only exceptions being the cases that start on unstable fixed points. Such fixed points reveal themselves by looking at the probability distributions of $T_{CW,n}$, $T_{CCW,n}$, and $R_{n}$,  denoted by $P(T_{CW,n})$, $P(T_{CCW,n})$, and $P(R_{n})$. For example, if $P(T_{CW})$, $P(T_{CCW})$, and $P(R)$ remain invariant under the block-scattering transformation, we know that we have reached a scale invariant point.

An attentive reader may object that such an averaging procedure performed on the RG scheme of Fig. \ref{fig:RGI} may be unpractical: since the size of the system shrinks at each iteration, the size of the set $\bigl\{\boldsymbol{S}_{n}^{\prime}\bigr\}$ depends on $n$, which is problematic to develop a rigorous statistical study of the flow over many iterations. At the same time, the numerical calculations of block networks and unitary matrix transformations increase exponentially with system's size, preventing us from working with very large systems. The problem of developing a consistent statistical description of the network and the one of computational efficiency are therefore intertwined.

We handle this issue by enhancing the RG scheme with a well-known strategy, which is able to successfully depict the effect of disorder especially in spin glass, and known as the replica scheme \cite{altland_anderson_2005,parisi_replica_1997,parisi_physical_2002,castellana_renormalization_2013,angelini_real_2017}. This way, we enforce the invariance of the size $P$ of the network during RG. In a nutshell, it is implemented by (i) making $L \gg P$ replicas of the network at the step $n-1$ (replicas in the sense that their scattering matrices  $\bigl\{\boldsymbol{S}_{n-1}^{\prime}\bigr\}$ have identical statistics, and the disorder statistics are also the same), (ii) concatenate all the network replicas into $U(3)$ matrices ($L$ members) by performing block scattering transformations, and finally (iii) use $P$ of these $U(3)$ matrices to construct at least $L$ new disordered networks that will form the replicas at the step $n$ (See Appendix \ref{ap:numerical}). One iteration therefore corresponds to an effective scaling of the network by a factor $P^{\prime}<P$, yet without changing the number of scatterers at each iteration. The size $P$ can be maintained relatively small, as long as accurate statistics can be performed, making it possible to look at arbitrarily large numbers of RG iterations at relatively low computational cost, and explore the physics occurring at scales with largely different orders of magnitude.

\subsection{Results for random phase-link disorder}
Taking the matrices $\boldsymbol{S}_1$ and $\boldsymbol{S}_2$ used in Fig. \ref{fig:finiteexamples} as examples, we use the replica scheme and go through seven RG iterations. The results are presented in Fig. \ref{fig:two examples}. Consistent with the finite-size simulations of Fig. \ref{fig:finiteexamples}, $\boldsymbol{S}_1$ and $\boldsymbol{S}_2$ are found to converge to different attractors:  $\boldsymbol{S}_R$ for $\boldsymbol{S}_1$ (Fig. 6a), indicating a trivial insulator, and $\boldsymbol{S}_{CW}$ for $\boldsymbol{S}_2$ (Fig. \ref{fig:two examples}(e)), indicating a topological insulator. To see how the macroscopic scattering properties change upon scaling up the disordered networks, we track the evolution of $P(T_{CW})$, $P(T_{CCW})$, and $P(R)$ during iterations. As shown in Fig. \ref{fig:two examples}(b) and (f), after five iterations the attractors are almost reached: although both systems are initially clockwise chiral, only disordered systems based on $\boldsymbol{S}_2$ maintain high values of $T_{CW}$ at large scales. Conversely, for large enough disordered networks based on $\boldsymbol{S}_1$, $T_{CW}$ and $T_{CCW}$ gradually disappear, while the probability of observing full reflection is close to unity after the fifth iteration. These distinct behaviors are further confirmed by the evolution of the averages $\langle T_{CW} \rangle$, $\langle T_{CCW} \rangle$, and $\langle R \rangle$, exhibited in Fig. \ref{fig:two examples}(c) and (g). A visual summary of this process is obtained by plotting the RG flows, starting from the initial point $\boldsymbol{S}_2$ ( $\boldsymbol{S}_1$) in the parameter space, and going through a discrete trajectory to reach $\boldsymbol{S}_{CW}$ ($\boldsymbol{S}_R$).  

\begin{figure}[htbp!]
\includegraphics[width=0.48\textwidth]{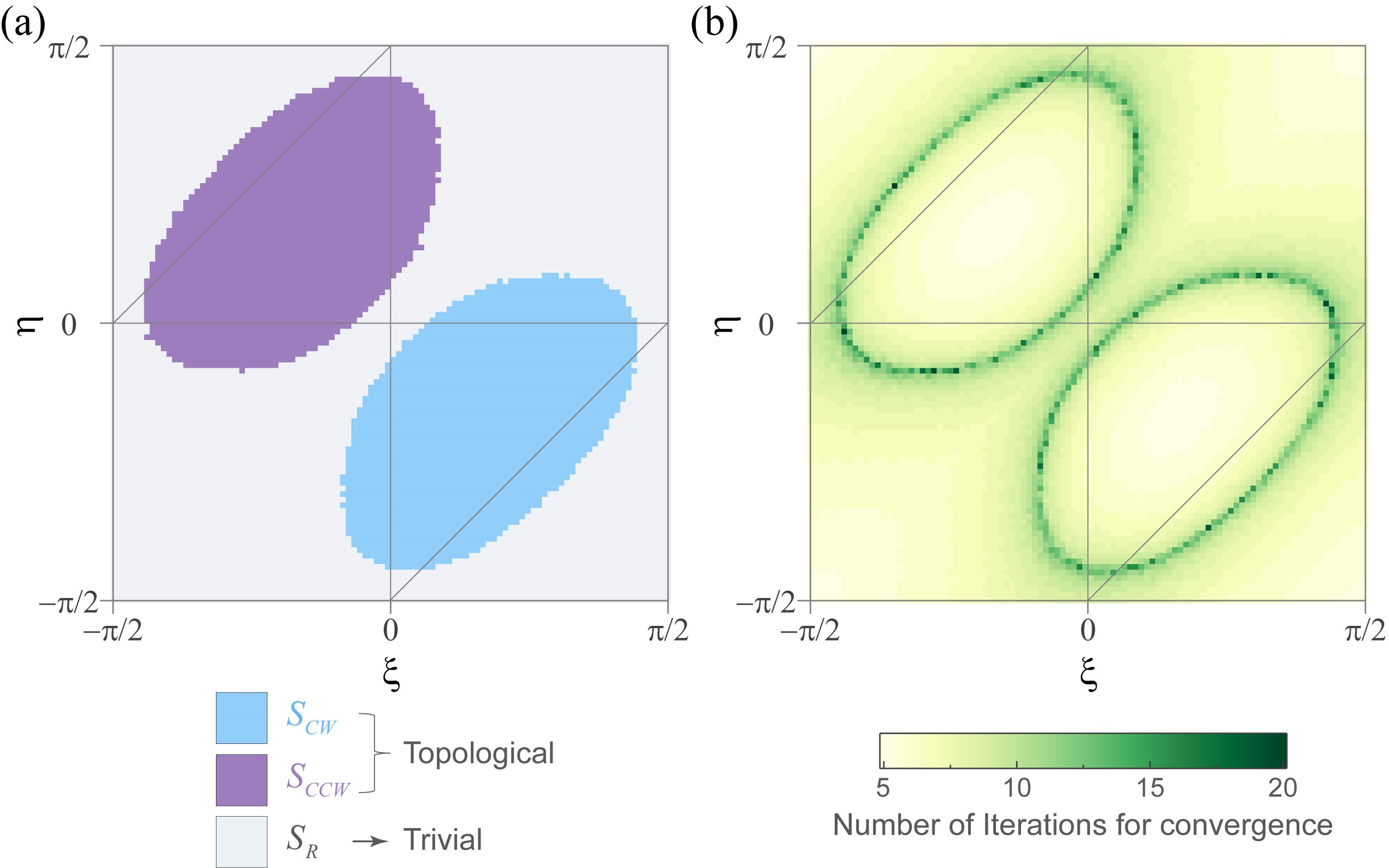}
\caption{ \textbf{RG phase diagram. }(a) Phase diagram obtained by summarizing, for each network built from $\boldsymbol{S}_0(\xi,\eta)$, which RG scattering attractor is reached after convergence of RG flow. (b) Number of RG iterations required to reach a fixed level of convergence to the scattering attractor. Local maxima are obtained at the phase transition, which is a signature of scaling invariant behaviors.}
\label{fig:RG_phase_digram}
\end{figure}
\begin{figure}[htbp]
\includegraphics[width=0.48\textwidth]{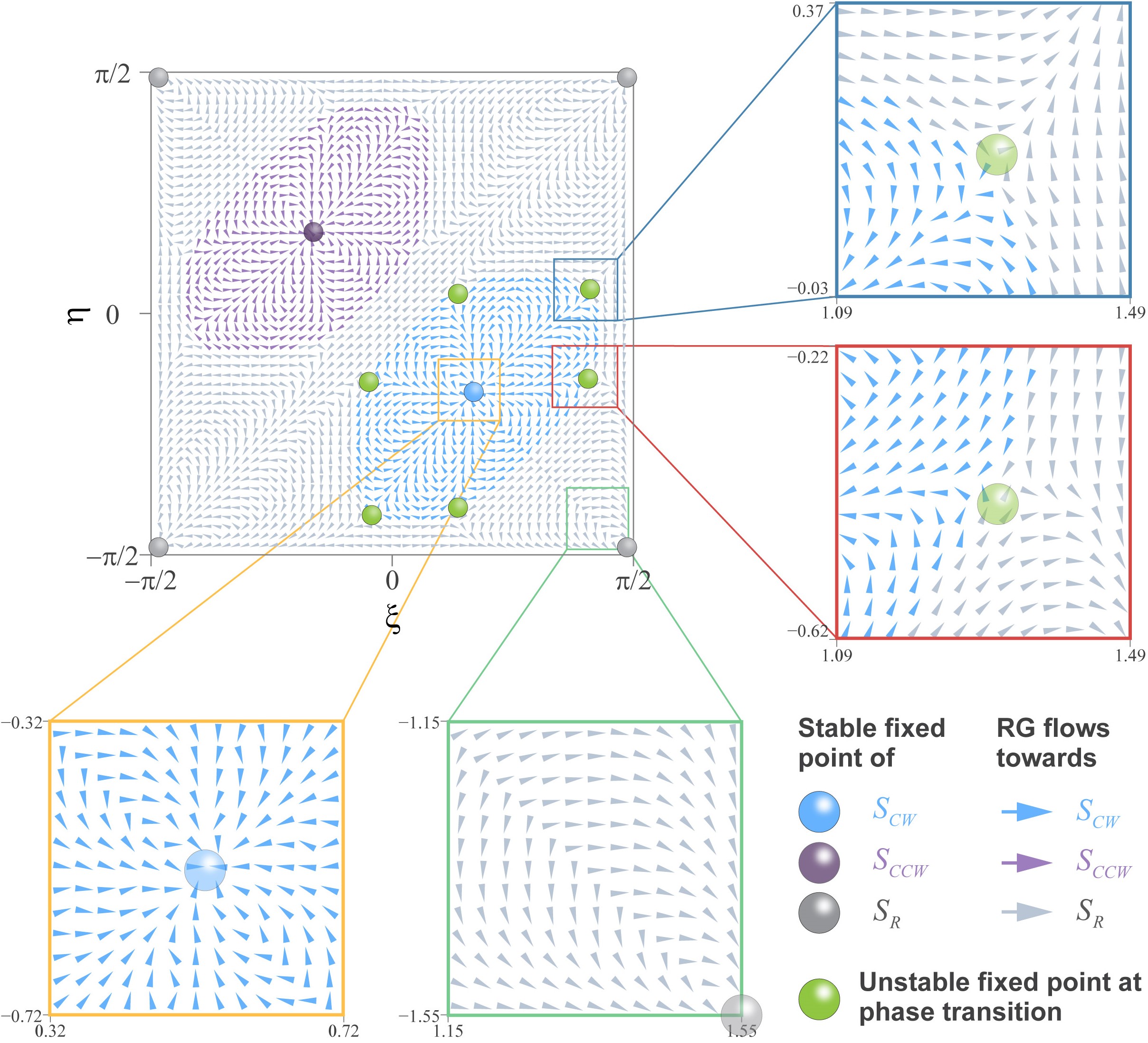}
\caption{ \textbf{RG flow diagram. }Arrows in the flow diagram indicate the effect of successive RG transformations on the macroscopic scattering properties of scaled networks. Non-zero flow divergence confirm that the points  $\boldsymbol{S}_{CW}$, $\boldsymbol{S}_{CCW}$, and $\boldsymbol{S}_R$ located at the centers of three phases are stable fixed points (a zoomed in view is provided by the two bottom insets). Six saddle points (green) define unstable fixed points located on the topological phase boundary.}
\label{fig:RG_flow_phase}
\end{figure}

This procedure, performed so far for only two initial scatterers $\boldsymbol{S}_1$ and $\boldsymbol{S}_2$, can be repeated for all possible initial scatterers that belong to the $(\xi,\eta)$ plane, obtaining a RG phase diagram for phase-link disorder. The RG phase diagram summarizes, for each $\boldsymbol{S}_0(\xi,\eta)$, the attractor reached by large disordered networks built from $\boldsymbol{S}_0$ (Fig. \ref{fig:RG_phase_digram}(a)). There are two topological phases of opposite chirality, depending on whether large systems converge to $\boldsymbol{S}_{CW}$ or $\boldsymbol{S}_{CCW}$ . They are separated by the trivial phase, composed of systems that converge towards the full-reflection attractor $\boldsymbol{S}_R$ upon scaling. At the interfaces between systems converging to $\boldsymbol{S}_{CW}$ (or $\boldsymbol{S}_{CCW}$), and systems converging to $\boldsymbol{S}_R$, a topological phase transition occurs, which is related to the presence of a critical metal with infinitely long field correlations. As a result, the RG approach requires more iterations around these transition lines before choosing an attractor, consistent with the observation of a extremum of the number of RG iterations required for convergence at the boundary, as shown in Fig. \ref{fig:RG_phase_digram}(b). 

To better represent the process of converging towards stable fixed points, we plot the associated RG flow diagram, shown in Fig. \ref{fig:RG_flow_phase}. Any initial value of $(\xi,\eta)$ in the blue (purple) region leads to a trajectory heading to the clockwise chiral attractor $\boldsymbol{S}_{CW}$ ($\boldsymbol{S}_{CCW}$). However, any point starting in the grey region goes to the fully reflective attractor $\boldsymbol{S}_{R}$.  The interface between the two regions are critical lines connecting six unstable fixed points (saddle points) of the RG flow. Any point with arbitrary small deviations from the critical lines will flow away from the critical condition, and eventually converge to a stable fixed point. In such disordered systems, a topological phase transition is equivalent to the crossing of a critical line, which represent systems that reach an exact balance between nonreciprocity and reflection upon scaling.

\begin{figure}[htbp]
\includegraphics[width=0.48\textwidth]{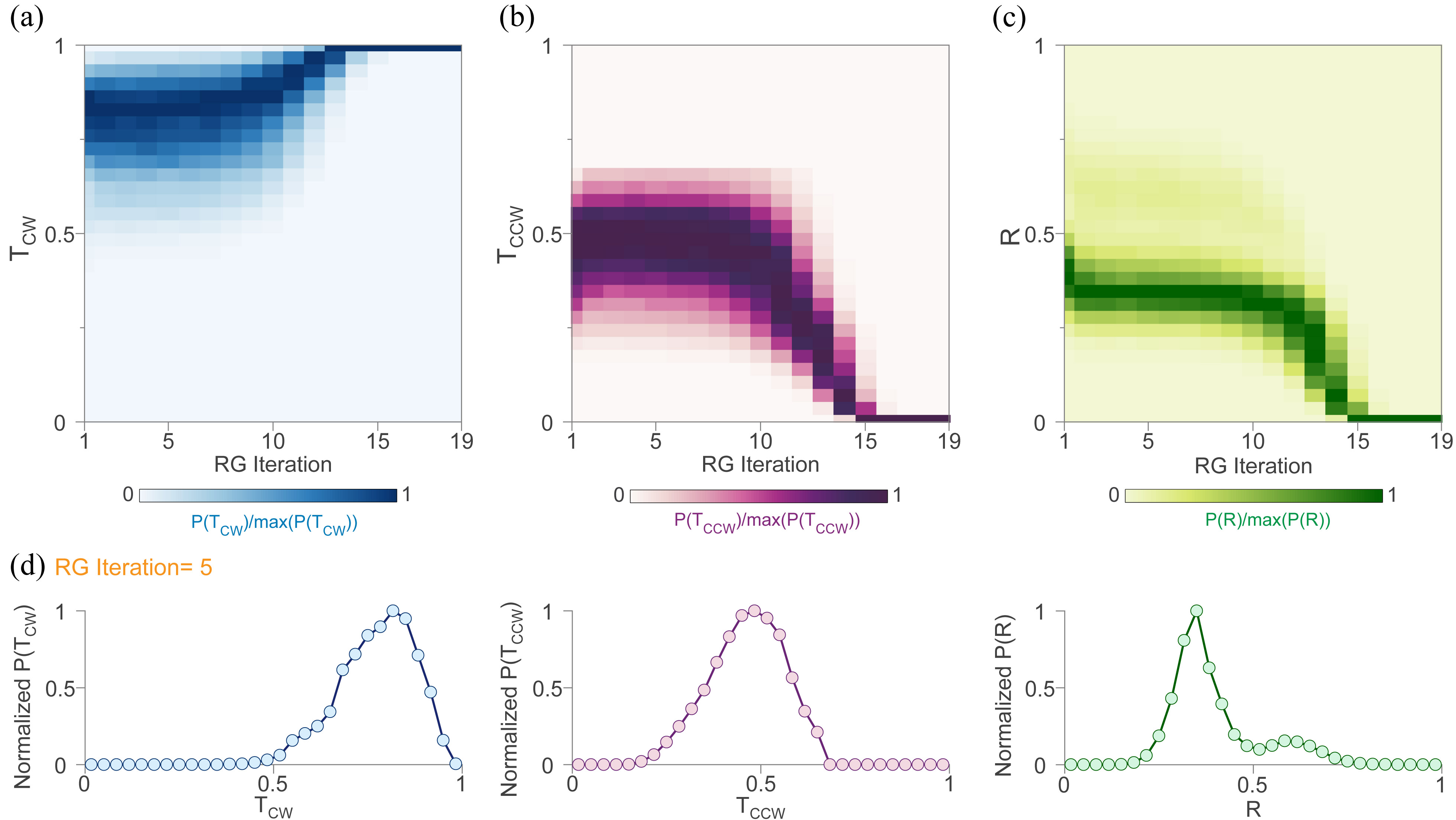}
\caption{ \textbf{Critical scattering properties. }(a-c) Evolution of $P(T_{CW})$, $P(T_{CCW})$, and $P(R)$ of the scattering matrix at the point  $\xi= -\eta= 0.92$, which is close to the critical point between $\boldsymbol{S}_1$ and $\boldsymbol{S}_2$, but on the side of $\boldsymbol{S}_{CW}$. The probability distributions of  $T_{CW}, T_{CCW}, R$ are invariant over the first ten RG scaling iterations, consistent with critical behavior. (d) Corresponding critical probability distributions at the topological phase transition in phase-disordered honeycomb scattering networks.}
\label{fig:probab_critical}
\end{figure}

To complete the picture, we explore RG as close as possible to the critical scattering matrix located on the line between $\boldsymbol{S}_1$ and $\boldsymbol{S}_2$ in the parameter space. For this, we choose a point very close to the critical scattering matrix $\boldsymbol{S}_c$, but on the side of $\boldsymbol{S}_{CW}$. We expect that such a point would behave similarly as $\boldsymbol{S}_c$ during RG, at least during the first iterations, before it starts moving toward the attractor. As shown in Fig. \ref{fig:probab_critical}, the probability distributions $P(T_{CW})$, $P(T_{CCW})$, and $P(R)$ are invariant in the first eleven iterations, which is indeed a symptom of the scale invariance expected for critical phenomena. Furthermore, the critical probability distributions of $T_{CW}, T_{CCW}$, and $R$ represented in Fig. \ref{fig:probab_critical}(d) establish a reference to determine whether a disordered network with arbitrary disorder will be topological or trivial in the thermodynamic limit. We emphasize that the above critical distributions for phase transitions between trivial and topological insulators describe a totally different physical situation from the metal- insulator transition of quantum Hall effect already captured by CC networks \cite{shapiro_renormalization-group_1982,galstyan_localization_1997,janssen_s-matrix_1998,klesse_spectral_1997,chalker_percolation_1988,metzler_influence_1999,cain_real-space_2003,kramer_random_2005}. For CC networks governed by quantum percolation, the phase diagram consists instead of two phases, metal and insulator, and is described by a single scalar quantity $T$. In contrast, disordered topological scattering networks are described by the distributions of the three quantities $T_{CW}, T_{CCW}, R$ required to describe the topological criticality, which would no be possible with a single scalar quantity $T$ that contains no information about the chirality of the transport.







%% file: SectionIV.tex
\label{SectionIV}
In this section, we confront the phase transition boundary obtained from RG in the previous section to the one obtained from a different, computationally more intensive method based on calculating the localization length. The agreement between these independent results confirm the quantitative accuracy of the RG scheme. As a by-product of the localization length study, we obtain insights on the topological phase diagram of phase-disordered networks and shed light on the associated critical phenomena. 

\subsection{Localization length and critical exponents}



Localization and correlation lengths play a central role in analyzing phase transitions and critical behaviors in disordered condensed matter systems \cite{merkt_network_1998,kramer_random_2005,onoda_localization_2007,obuse_two-dimensional_2007,liu_effect_2016,liu_supermetal-insulator_2021,luo_unifying_2022,xiao_anisotropic_2023,wang_anderson_2023} and photonics \cite{schwartz_transport_2007,rotter_light_2017,yu_engineered_2021,vynck_light_2021,yamilov_anderson_2023}. Benefiting from the direct availability of transfer matrices, network models are particularly suited to quantitative scaling analysis and the description of phase transitions. Early arts focused on the scaling theory of localization \cite{anderson_new_1980,mackinnon_mackinnon_1982} and the study of metal-insulator transitions in QHE \cite{chalker_percolation_1988,slevin_critical_2009,amado_numerical_2011}, exhibiting precise critical exponents. In recent years, these networks have been widely applied on disordered topological systems, to quantitatively characterize topological phase boundaries and their universality, including topological systems protected by time reversal symmetry \cite{obuse_two-dimensional_2007,obuse_boundary_2008,obuse_spin-directed_2014}, in 2D or 3D \cite{son_3d_2021}. In this section, we apply the transfer matrix method to study localization processes and topological phase transitions in unitary scattering networks with random phase-link disorder. 

\begin{figure}[htbp]
\includegraphics[width=0.48\textwidth]{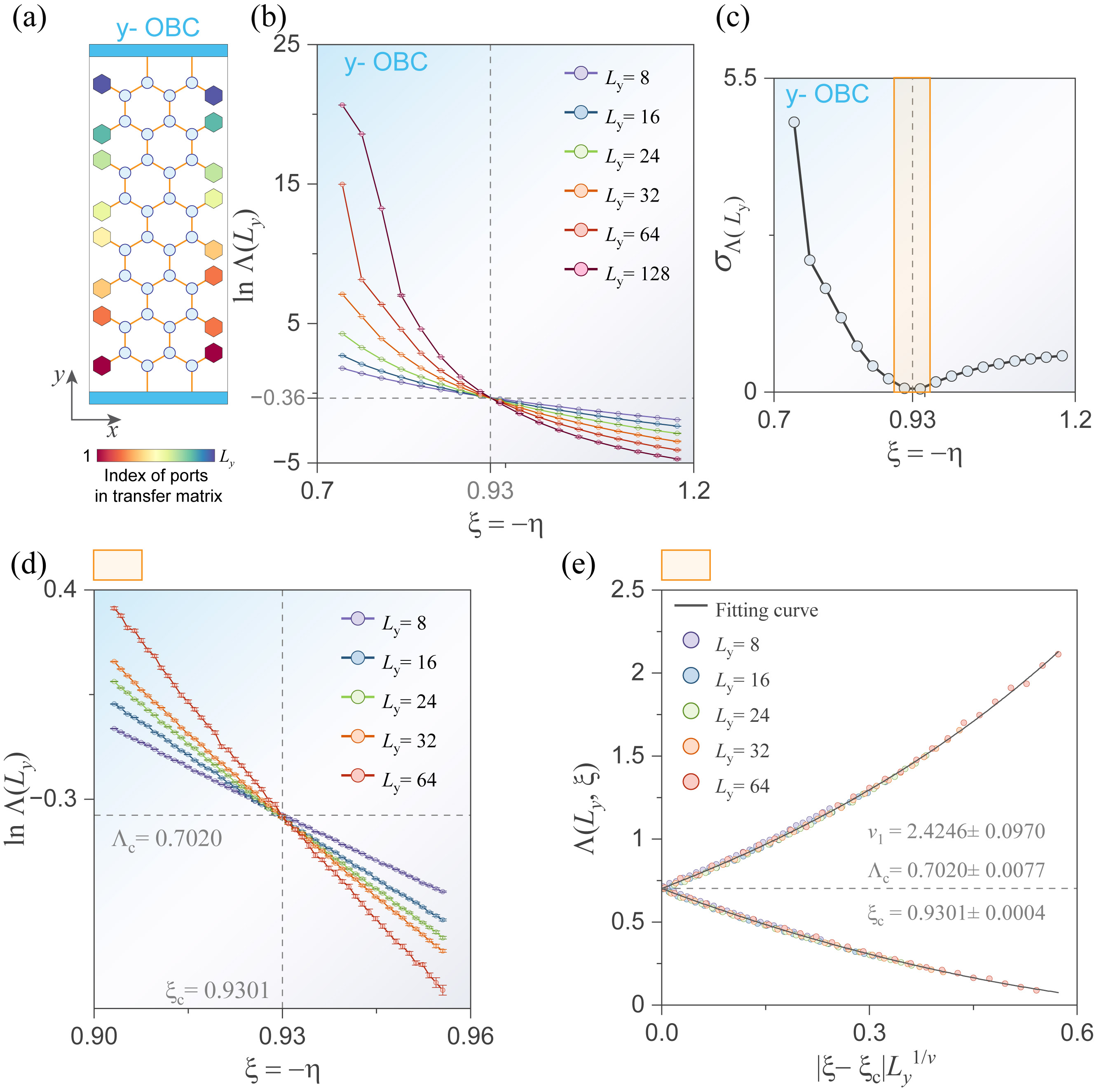}
\caption{ \textbf{Topological phase transitions and critical behaviors by a scaling analysis of the localization length.} (a) A slice of a quasi-1D network with  width $L_y= 8$ is used for iteratively calculating the transfer matrix. The longitudinal dimension contains four elementary slices, therefore $L_x= 4$. (b) Evolution of the normalized localization length $\Lambda(L_y)=\lambda/L_y$ with the width $L_y$ on the segment of $\xi=-\eta \in [0.73, 1.18]$. $L_y$ is increased from 8 to 128. Error bars are smaller than the markers. The left and right parts of the plot are topological and trivial phases, respectively, and are separated by the critical point $\xi_c \approx 0.93$ (dashed lines) characterized by scale invariance. (c) Criticality can be numerically identified as the local minimum of the standard deviation of the normalized localization length, $\sigma_{\Lambda(L_y)}$.(d) Zoomed-in results of $\Lambda(L_y)$ scaling in the vicinity of $\xi=-\eta= \xi_c$ (marked as yellow region in panel (c)). The dashed lines mark the critical point $\xi=-\eta= \xi_c=0.9301$ and $\Lambda_c= 0.7020$. (e) Single parameter scaling function $\Lambda$ fitted from the data in (d). We obtain a critical exponent $\nu_1= 2.4246 \pm 0.0970$. The error bars correspond to the $95\%$ confidence intervals estimated from the Monte Carlo simulations.}
\label{fig:LL_093}
\end{figure}

To derive the localization length, we work on a quasi one-dimensional (1D) network, whose total longitudinal size $M_x$ is much larger than its transverse size $L_y$. Both lengths are defined by counting the number of ports on the $x$ and $y$ directions, respectively. The number of ports on any opposite sides are equal. We look at the properties of the transfer matrix $T$ connecting the fields of the ports on the right boundary to the fields of the ports on the left boundary, i.e. in the $x$ direction. $T$ is then a $2L_y \times 2L_y$ pseudo unitary matrix \cite{munshi_self-adjoint_2019}, defined by the scattering relations among the $2L_y$ ports located on on the lateral sides. Due to the pseudo unitary of $T$, the eigenvalues of the $2L_y \times 2L_y$ Hermitian matrix $T^\dag T$ can be written as $\text{exp}(\pm 2X_j)$ with eigenstates $|\Psi_j^{\pm}\rangle$, where $X_j$ are Lyapunov exponents such that $0<X_1<X_2< \cdots <X_{L_y}$. As the wave transport counted over $2L_y$ eigenchannels is dominated by the contribution of the smallest Lyapunov exponent $X_1$ ($\langle\Psi_1^{-}| T^{\dag}T|\Psi_1^{-}\rangle$), the localization length $\lambda$ is defined as the inverse of the smallest Lyapunov exponent \cite{mackinnon_mackinnon_1982,merkt_network_1998,kramer_random_2005} as
\begin{equation}
    \lambda=\lim_{M_x\to\infty}\frac{M_x}{X_1}.
    \label{eq:lamda}
\end{equation}
Calculating and decomposing $T$ leads to large numerical errors when $M_x$ is large ($> 20$), yet $M_x$ should be at least millions to ensure $M_x> \lambda$.  To reduce numerical errors in the determination of the localization length, we uniformly slice the long quasi-1D network along the $x$ direction, with a slice width $L_x$, and get slices indexed from $i=1$ to $M_x/L_x$. We calculate the transfer matrix $T_i$ of each slice, and then multiply them by iterative QR decomposition \cite{slevin_critical_2014}:
\begin{equation}
\begin{cases}
    T= \displaystyle \prod_{i}^{M_x/{L_x}}  T_{i};\\
    T_1=Q_1R_1;\\
    T_{i+1}Q_i=Q_{i+1}R_{i+1}, & i>1.
\end{cases}
\end{equation}
Therefore, we get 
\begin{equation}
     T= Q_{M_x/{L_x}} \left[ \prod_{i}^{M_x/{L_x}} R_{i}\right] \equiv Q_{T}R_{T},
\end{equation}
where $R_{T}\equiv\prod_{i}^{M_x/{L_x}}R_{i}$ and $Q_{T} \equiv Q_{M_x/{L_x}}.$ As Lyapunov exponents are exactly the diagonal elements of the upper triangular matrix $R_{T}$, we can therefore get 
\begin{equation}
    \lambda=\lim_{M_x\to\infty}\frac{M_x}{\displaystyle \min_j |\text{ln }R_{T}(j,j)|}.
     \label{eq:lamda2}
\end{equation}
In this way, we maintain a good accuracy on the eigenvalue with modulus closest to unity. Generally, the smaller $L_x$ is, the more accurate $T$ is. Here, we take $L_x \in \{4,8,12,16\}$. In Fig. \ref{fig:LL_093}(a), we show a slice of such a system of length $L_x=4$  ($L_x\ll M_x$ ), and of transverse size $L_y= 8$. We consider open boundary conditions (OBC) in the $y$-direction, namely edges with unitary reflection. As indicated by Eq. \ref{eq:lamda2}, calculating the exact localization length $\lambda$ requires taking $M_x$ to infinity. Also, in disordered networks, $T_i$ is taken from a statistical ensemble of finite quasi-1D network slices which are composed of specified microscopic scatterers and under prescribed disorder statistics. Therefore, it seems that one should take $M_x$ very large and average $\lambda$ over many calculations. Fortunately, as localization length $\lambda$ is a finite and self-averaging quantity \cite{slevin_critical_2014}, we can approximate $\lambda$ by only one calculation on a finite, but long enough, quasi-1D network with $M_x \in [5 \times 10^5, 2\times10^6]$.



One can identify whether a system is in an insulating or metallic phase by analysing how the normalized localization length  $\Lambda(L_y)=\lambda/L_y$ scales when increasing the width $L_y$, as demonstrated in prior arts \cite{mackinnon_one-parameter_1981}. This is traditionally done by checking the dependence of $\Lambda(L_y)$ on the transverse width $L_y$, when applying periodic boundary condition (PBC) in the $y$ direction (thus in a setting with no top and bottom edges). For a metal, $\Lambda(L_y)$ increases with $L_y$, and $\Lambda(L_y) \to \infty$ as $L_y \to \infty$. On the contrary, for an insulator, $\Lambda(L_y)$ decreases upon scaling $L_y$, and $\Lambda(L_y) \to 0$ as $L_y \to \infty$. At a critical transition, $\Lambda$ should be invariant upon scaling. This method, however, is not sufficient to distinguish topological and trivial insulators. To this end, one should repeat the study in the presence of edges, namely with PBC replaced by OBC \cite{halperin_quantized_1982}. Since nontrivial topology manifests itself by the existence of chiral edge states, $\Lambda(L_y)$ should increase monotonically, like for a metal. On the other hand, the behavior of a topologically trivial insulating phase would be insensitive to the modification of the boundary condition \cite{halperin_quantized_1982}, due to the absence of edge states.

In Fig. \ref{fig:LL_093}(b), we show the results of such scaling analysis, with OBC applied in the $y$ direction (Fig. \ref{fig:LL_093}(a)). We vary the network parameters along the line $\xi= -\eta$, ranging from $0.73$ to $1.18$, expecting to cross the phase transition. We observe a monotonically increasing $\Lambda(L_y)$ in the range $0.73< \xi< \xi_c \approx 0.93$, where $\xi_c$ denotes the critical value for which $\Lambda(L_y)$ is scale-invariant, marked by a dashed line. Conversely, $\xi_c < \xi< 1.2$ exhibits an insulator behavior. To determine whether the range $0.73< \xi< \xi_c$ corresponds to a metal or a topological insulator, we changed the boundary conditions to PBC,  which reversed the scaling behavior of the localization length, excluding the metal (See Appendix \ref{app:LL}, Fig. \ref{fig:PBC_LL}). Therefore, we conclude that disordered networks with $0.73< \xi< \xi_c$ are topological insulators, whereas $\xi_c < \xi< 1.2$ are trivial insulators. Numerically, the topological criticality at $\xi_c \approx 0.93$ can be identified by a local minimum in the standard deviation of $\Lambda(L_y)$ for a set of $L_y$, represented as $\sigma_{\Lambda(L_y)}$ in Fig. \ref{fig:LL_093}(c). 



The behaviors near phase transition are known to follow power laws, where the critical exponent $\nu$ is of prime interest to characterize their universality class. To extract the critical exponent $\nu$, we focus on a smaller range around the transition, shaded in orange in Fig. \ref{fig:LL_093}(c). The scaling of $\Lambda$ in the vicinity of the critical point $\xi= \xi_c \approx 0.93$ is shown in Fig. \ref{fig:LL_093}(d). If the localization length $\lambda$ diverges following the power law
\begin{equation}
    \lambda \sim |\xi-\xi_c|^{-\nu},
\end{equation}
with the single parameter scaling ansatz, the normalized localization length $\Lambda(L_y,\xi)$ can be expanded as
\begin{equation}
    \Lambda=\Lambda_c + \sum_{q=1}^{Q_1} a_{q} \left[ (\xi-\xi_c) L_y^{\frac{1}{v}}\right] ^{q}+ \sum_{q=0}^{Q_2} b_{q} \left[ (\xi-\xi_c) L_y^{\frac{1}{\nu}}\right]^{q} L_y^{z},
\label{Eq:scaling ansatz}
\end{equation}
where the third term is a finite-size effect correction \cite{slevin_corrections_1999}, with $z$ defined as a negative exponent. Therefore, in the limit of $L_y \to \infty$, Eq. (\ref{Eq:scaling ansatz}) recovers the standard form of single-parameter scaling as 
\begin{equation}
    \lim_{L_y\to\infty}\Lambda :=\Lambda_c + \sum_{q=1}^{Q_1} a_{q} \left[ (\xi-\xi_c) L_y^{\frac{1}{\nu}}\right] ^{q}.
\label{Eq:one para scaling }
\end{equation}
With the help of Eq. (\ref{Eq:scaling ansatz}) and using $Q_1=5$ and $Q_2=2$, we fit the data in Fig. \ref{fig:LL_093}(d) and obtain all the parameters $\nu, \xi_c,\Lambda_c, z$, as well as the coefficients $a_1, a_2, a_3, a_4, a_5, b_0, b_1, b_2$. To reduce the statistical error, we averaged the result over 100 disorder realizations. Fig. \ref{fig:LL_093}(e) plots the single parameter scaling function $\Lambda(L_y,\xi)$ (solid line) together with the data (dots) as a function of $L_y$. Our estimation lead to the critical exponent $\nu_1= 2.4246 \pm 0.0970$ and critical length $\Lambda_c= 0.7020 \pm 0.0077$ for the critical boundary at $\xi= -\eta \approx 0.9301$. We observe that the value of $\nu_1$ and and $\Lambda_c$ are very close to the ones reported for QHE \cite{slevin_critical_2009}, which is not surprising since they are both topological insulators belonging to class A \cite{slevin_critical_2009,amado_numerical_2011,luo_unifying_2022}. 
\subsection{Topological phase diagrams comparisons and two critical exponent values}
\begin{figure}[htbp]
\includegraphics[width=0.48\textwidth]{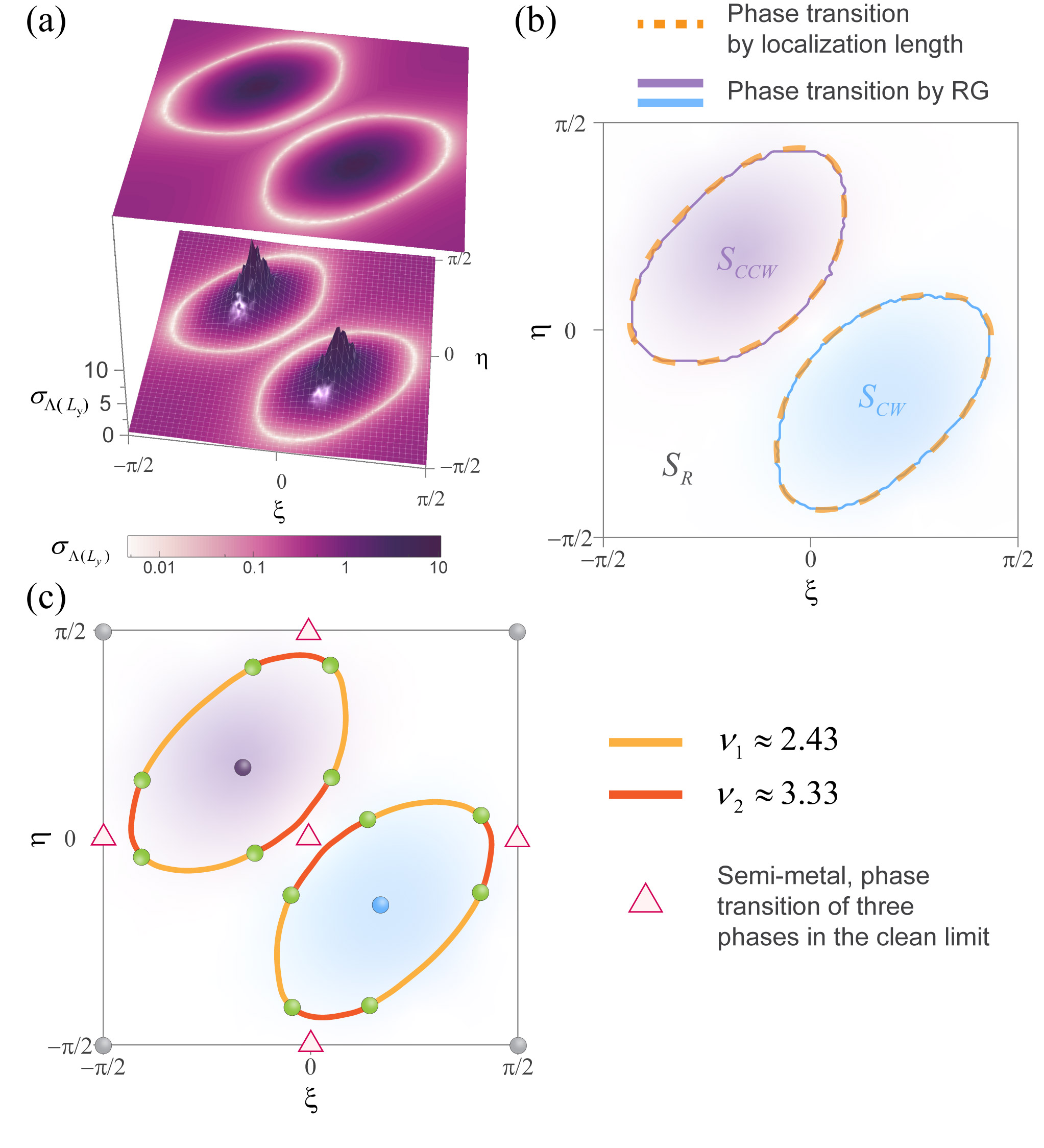}
\caption{ \textbf{Comparisons of the topological phase diagrams and critical behaviors obtained from RG and localization length (LL) analysis.} (a) In LL analysis, the boundaries of topological phases are revealed by the local minima of $\sigma_{\Lambda(L_y)}$. (b) Direct comparison of the RG and LL topological phase diagrams. They agree with percent-level accuracy. (c) Critical exponent distribution on the phase boundaries. Two values of critical exponents - $\approx 2.43$ (orange) and $\approx 3.33$ (red)- emerge along the critical boundary, and change only at the RG unstable fixed points, which are saddle points of the RG flow.}
\label{fig:comp_RG_LL}
\end{figure}

The study of the previous section can be repeated for any point of the phase diagram, in order to directly confront the topological boundary predicted by RG to the one obtained from the scaling analysis of the localization length. The later can be obtained by looking at the location of local minima of $\sigma_{\Lambda(L_y)}$ in the parameter space, plotted in Fig. \ref{fig:comp_RG_LL}(a). In Fig. \ref{fig:comp_RG_LL}(b), we show the direct comparison of the phase boundaries obtained from localization length calculations (dashed line) and from RG (solid lines). We observe a very good agreement between the two, validating the validity of our RG scheme, with percent-level quantitative accuracy. In terms of numerical efficiency, the time cost of RG calculation for one point of the phase diagrams is a few seconds, (minutes very close to critical points), versus several hours for the corresponding localization length calculations, performed on Intel(R) Xeon(R) Platinum 8360Y processors.

Surprisingly, we found that the critical behaviors vary discretely on the topological phase boundaries. More specifically, they are of two kinds, and change at the RG flow saddle points. The values of the two distinct critical exponents are $\nu_1 \approx 2.43$ and $\nu_2 \approx 3.33$, and their locations are shown in Fig. \ref{fig:comp_RG_LL}(c). A detailed example for $\nu_2 \approx 3.33$ in the vicinity of $\xi= -\eta  \approx 0.0947$ is shown in Appendix \ref{app:LL}, Fig. \ref{fig:LL_035}, which is also found on the segment $\xi= -\eta \in [0, \pi]$,  on the other side of the phase diagram. A given segment connecting two unstable fixed points (saddle points) in the RG flow diagram exhibits a constant critical exponent value, which changes when crossing saddle points. We conjecture that such distinct critical behaviors nucleate from distinct types of topological phase transitions, the origin of which can be clearly seen on the clean-limit topological phase diagram. The critical segments with $v_1$ are associated to the topological phase transition between the anomalous and Chern phases in the clean limit (Appendix \ref{ap:cleanlimit}, Fig. \ref{fig:cleanPhasediagram}), whereas the ones with $v_2$ can be attributed to more complex phases transitions at the semi-metal points in the clean limit, where three topological phases meet: anomalous, Chern, and trivial insulators.




%% file: SectionV.tex
\label{SectionV}
\begin{figure*}[htbp]
\includegraphics[width=1\textwidth]{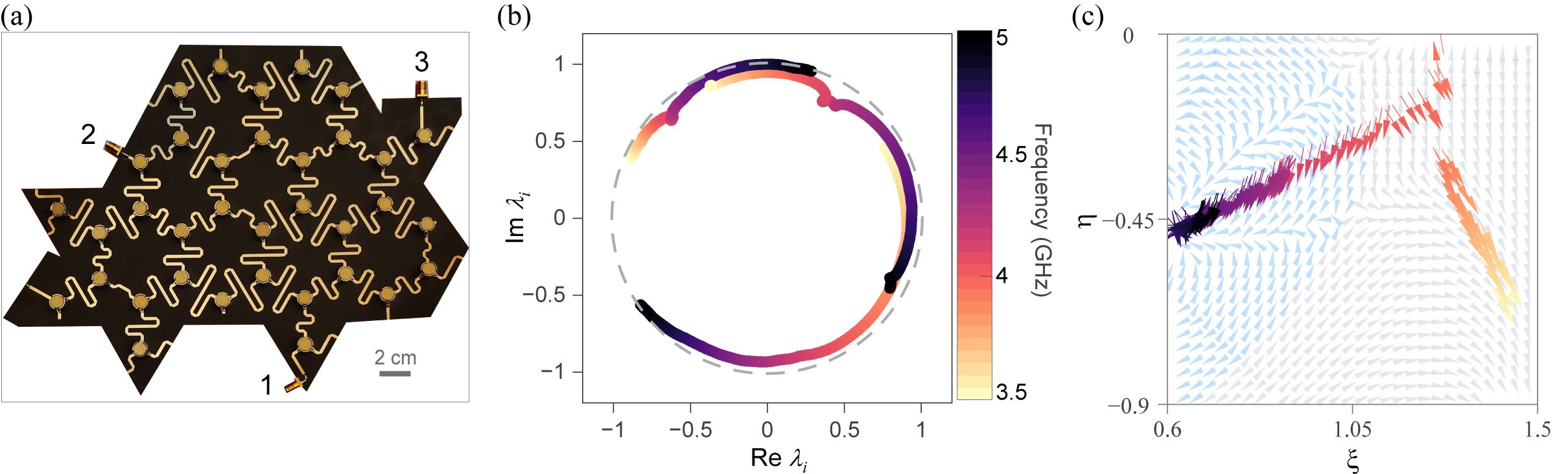}
\caption{\textbf{Experimental validation of renormalization group flow.} (a) Photograph of one of our 5 prototypes. We experimentally validate the scattering RG flow by using microwave networks made of ferrite circulators interconnected by microstrip lines. The ferrite circulators serve as microscopic three-port scatterers, and their scattering matrix depends continuously on frequency, allowing us to access a continuous family of networks with different attractors simply by varying the frequency of operation.  The random phase-link disorder is realized by varying the length of the meandering microstrips that connect the circulators together. The 5 prototypes correspond to different realizations of disorder. At each frequency, we can measure the microscopic scattering properties of a single circulator, as well as the macroscopic scattering properties of the networks, taken at three external probes located on the boundary. This allows us to experimentally extract the RG flow. (b) Experimentally measured eigenvalues of the scattering matrix of the circulators, confirming the quasi-unitarity of the microscopic scattering process over the experimental frequency band ($3.5- 5$ GHz). (c) Measured RG flow (colored arrows) averaged over the five different disorder realizations. The arrows composing the background are numerical predictions of the RG flow ($M_i=1, i=1,2,3$). Each color corresponds to a particular frequency.}
\label{fig:exp}
\end{figure*}
We performed experiments with microwaves in scattering networks with random phase-link disorder, built using off-the-shelf ferrite circulators. The disordered phase-delay links with $\Delta\varphi= 2\pi$ are experimentally achieved by serpentine microstrip lines, whose phase delay $\varphi$ under a length $L$ at the frequency $f$ is expressed as $\varphi= \frac{2\pi f L  \sqrt{\epsilon_{eff}}}{c}$, with $\epsilon_{eff}$ being the effective permittivity of the microstrip, obtained from standard microwave design formulas. We built five prototypes with different disorder realizations, and show a picture of one of them in Fig. \ref{fig:exp}(a). In the frequency range of interest, $f \in [3, 5]$ GHz, these circulators are nearly identical, and the experimentally measured scattering matrix $\boldsymbol{S}_0(f)$ is well approximated by a $C_3$-symmetric unitary matrix (Fig. \ref{fig:exp}(b)). $\boldsymbol{S}_0(f)$ therefore corresponds to a point on the $(\xi, \eta)$ parameter plane, whose precise location depends on frequency. At each frequency, we can not only measure the microscopic scattering properties of a single circulator $\boldsymbol{S}_0(f)$, but also the macroscopic scattering properties of the networks $\boldsymbol{S}_1^{\prime}(f)$, taken at three external probes located on the boundary. The difference between $\boldsymbol{S}_0(f)$ and $\boldsymbol{S}_1^{\prime}(f)$ is the experimental RG flow, shown by coloured arrows in Fig. \ref{fig:exp}(c). As the measurements of networks involve one probe per side, we compare the measured flow with the numerical RG flow obtained for $M_i=1, i=1,2,3$, shown by smaller arrows in the background. Clearly, the measured RG flow in the blue region points toward the center of blue region, namely the $\boldsymbol{S}_{CW}$ attractor, consistent with the numerical RG predictions. On the contrary, the flow measured in the grey region heads to the fixed point of $\boldsymbol{S}_{R}$, as expected from theory. Slight discrepancies between measured and predicted flows are observed nearby the critical boundary. They are attributed to the limited number of disorder realizations, as well as the limited size of the networks, which is not large enough to capture accurately the thermodynamic limit near critical boundaries. Nevertheless, our experiments substantiate the predicted RG flow and the convergence of large networks towards scattering attractors by confirming experimentally the accuracy of block-scattering transformations. We are able to confirm that the presence of phase-link disorder enhances the chirality of the transport when comparing $\boldsymbol{S}_1^{\prime}(f)$ to $\boldsymbol{S}_0(f)$, when the networks fall in the region of attraction of $\boldsymbol{S}_{CW}$ . Such results shed light on the origin of  topological chiral edge states in samples with strong distributed disorder.

%% file: SectionVI.tex
\label{SectionVI}
\begin{figure*}[htbp]
\includegraphics[width=0.8\textwidth]{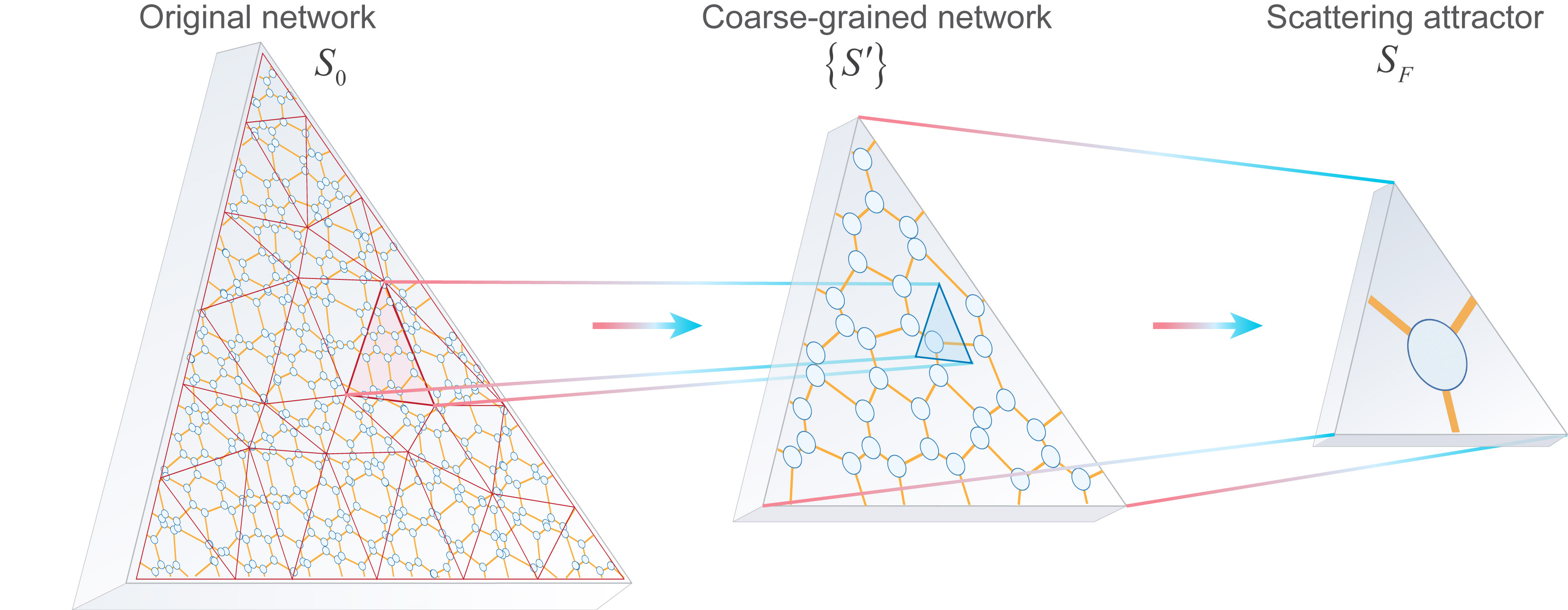}
\caption{\label{fig:RGII}\textbf{Renormalization group of a unitary scattering network with structural disorder.} As for the case of phase-disorder, the block-scattering transformation follows the three key steps of block division, transformation and reconstruction. The only difference is that the Delaunay triangulation is no longer regular. The transformation is used iteratively until one obtains a $U(3)$ attractor $\boldsymbol{S}_F$. In practice, this scheme is implemented via the replica scheme described in the section \ref{sec:Replica}}.
\label{fig:amor_RG_scheme}
\end{figure*}

\begin{figure*}[htbp]
  \centering
\includegraphics[width=0.8\textwidth]{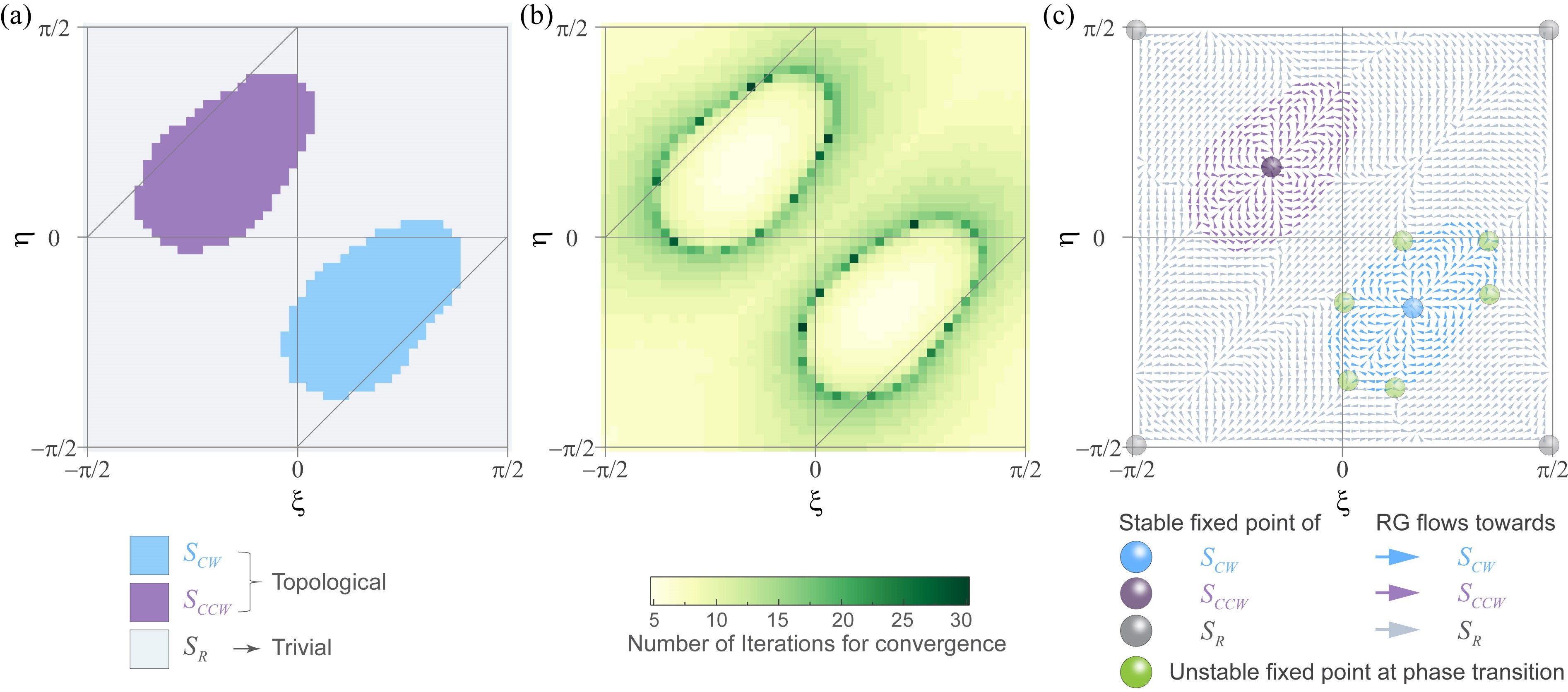}
\caption{ \textbf{RG topological phase diagram and flow diagram for networks under the strongest structural disorder.} (a) Topological phase diagram obtained from RG, representing the RG scattering attractor $\boldsymbol{S}_F$ for all possible choices for the microscopic scattering matrix in the parameter space. (b) Corresponding number of RG iterations required to converge to the attractor. (c) RG flow diagram, showing the transformation of networks upon iterative application of RG.}
\label{fig:RG_map_amor}
\end{figure*}
In the past several years, various works have demonstrated amorphous systems with nontrivial topology \cite{zhang_anomalous_2023,agarwala_topological_2017,mitchell_amorphous_2018,marsal_topological_2020,cassella_exact_2023,corbae_observation_2023}. This includes amorphous scattering networks supporting chiral topological edge states, which we recently observed experimentally \cite{zhang_anomalous_2023}. 

In this section, we therefore turn our attention to another type of disordered scattering networks with structural disorder. We implement our RG scheme to explore topological phase diagram and RG flows in such amorphous scattering networks. Technically speaking, these networks are generated by implementing a weighted Voronoi tessellation, with random weights on a triangular generator set lattice. This tessellation is the dual graph of the Delaunay triangulation. Its level of amorphism can be quantitatively described by the structural disorder factor $\alpha$, defined as the standard deviation of the random weights. The factor $\alpha$ is able to describe the continuous deformation of networks from pristine honeycomb ($\alpha= 0$) to the strongest possible level ($\alpha= 8$). In the following, we focus on scattering networks at the strongest level of structural disorder $\alpha =8$, although any other value of $\alpha$  can be chosen. More information about network statistics as a function of $\alpha$ can be found in recent works \cite{zhang_anomalous_2023,zhang_topological_nodate}.

The RG of amorphous networks is essentially similar to the one of phase-disordered honeycomb lattices. The implementation of the RG scheme on structurally disordered networks still relies on the same three steps (block division, transformation, and reconstruction). However, the Delaunay triangulation yielding the block division comes with a twist. Since we are dealing with networks with structural disorder at level $\alpha$, the Delaunay triangulation is no longer regular, but also at level $\alpha$ (Fig. \ref{fig:amor_RG_scheme}, leftmost). Following this, the dual graph that is used to interconnect the newly generated scatterers into a coarse-grained network (Fig. \ref{fig:amor_RG_scheme}, center) is also at disorder level $\alpha$.

The topological phase diagram obtained from RG is shown in Fig. \ref{fig:RG_map_amor}(a). It exhibits slightly smaller topological phase regions than the phase diagram observed in honeycomb networks with random phase-link disorder (Fig. \ref{fig:RG_phase_digram}). The phase boundary is confirmed by looking at the number of RG iterations needed for convergence into $\boldsymbol{S}_F$ (Fig. \ref{fig:RG_map_amor}(b)), whose local maxima indicates the scale invariance of networks located at the phase transition. Albeit with a slightly different topological range, the structure of the RG flow is very similar to the one found in the previous section, and exhibits the same landscape of stable and unstable fixed points as in Fig. \ref{fig:RG_flow_phase}, highlighting some for of critical universality between the two different kinds of disordered networks.




%% file: Conclusion.tex
\label{sec:conclusion}
We have presented a real-space renormalization group (RG) theory for unitary scattering network models, which offers significant insights into the emergence of topological edge states in large systems with strong distributed disorder. The method can reduce the scattering processes occurring in any unitary network into a 3x3 unitary scattering attractor that summarizes the key macroscopic scattering properties emerging at large scales. By introducing the block-scattering transformation, we focus on preserving key information about transport chirality and reflection, smearing out microscopic fluctuations into a macroscopic description of the scattering process. The combination of block scattering transformations and the replica strategy was shown to lead to a numerically-efficient RG scheme, capable of handling simultaneously an arbitrary number of RG iterations while performing Monte-Carlo simulations on disorder realizations. Our RG scheme is capable of discerning between topological and trivial disordered networks, since they correspond to distinct scattering attractors. Its implementation on two types of disordered networks not only clarifies the necessary microscopic conditions for constructing macroscopic topological networks, but also uncovers the unique critical phenomena occurring at topological phase boundaries, as well as the physics of disorder-resilient chiral edge transport. In addition, the quantitative accuracy of our RG framework is demonstrated through an independent scaling analysis of the localization length of quasi-1D networks, which predicts the same topological phase diagram as the RG scheme, with percent-level accuracy. As a by-product, we were able to elucidate the intricate critical phenomena occurring at the transition between disordered topological and trivial insulators, with critical exponents that take discrete values on the boundary, only changing at the RG flow saddle points. Finally, these theoretical advances are complemented by experimental verifications performed on microwave scattering networks, which are consistent with the calculated RG flow. We believe that such a scattering-based RG is general, and largely enhances a theoretical toolbox that may find direct applications in networks models used in condensed-matter and disordered topological physics. It also provides a computationally efficient method to predict the robustness of topological edge states in disordered systems. On the longer term, we envision that the block scattering transformation introduced in this work may also be useful in the study and understanding of large network models in communication systems, or in the development of physical neural-like networks based on scattering. With our methodologies and numerical frame, a potentially interesting future direction may be to explore other complex systems in the spirit of scattering, by grid discretization into networks or even by developing a continuous form of scattering RG that would address physical systems beyond networks. This could potentially unravel new aspects of the intricate interplay between topology, disorder and scaling. Practically, this research paves the way for designing more resilient and versatile devices in photonic, electromagnetic, and quantum computing networks, by establishing disorder as a general degree of freedom instead as a hindrance in the management and design of topological properties.

%% file: Appendix.tex
\label{Appendix}

\section{\label{ap:replacement} Recovery of one $\boldsymbol{S}^{\prime}\in U(3)$ from 3 by 3 non-negative matrix $\boldsymbol{E}_S$ }
As seen in the main text, when transforming $\boldsymbol{S}\in U(M)$ to $\boldsymbol{S}^{\prime}\in U(3)$, a key step is the recovery of a unitary matrix $\boldsymbol{S}^{\prime}$ from the non-negative matrix $\boldsymbol{E}_S$. As indicated by the recovery of the quark-mixing matrix from experimental data \cite{chau_comments_1984,dita_separation_2006}, a prerequisite for the $U(3)$ matrix recovery is to have a double stochastic matrix $\boldsymbol{A}_{DS}$. Double stochasticity is defined as
\begin{equation}
    \boldsymbol{A}_{DS}(i,j) \geq 0, \sum_{i=1}^{3} \boldsymbol{A}_{DS}(i,j)=1, \sum_{j=1}^{3} \boldsymbol{A}_{DS}(i,j)=1,
    \label{Eq:A_DS condition}
\end{equation}
which forms the Birkhoff polytope. The matrix  $\boldsymbol{A}_U$ with $\boldsymbol{A}_U(i,j):= \left| \boldsymbol{S}^{\prime}(i,j) \right|^2$, defined as the energy part of a unitary matrix $\boldsymbol{S}^{\prime}$, belongs to the set of double stochastic matrices $\{\boldsymbol{A}_{DS} \}$.   As a result, our recovery contains two steps. Firstly, with $\boldsymbol{E}_S$ in Eq. (\ref{eq:ES}), we obtain the corresponding $\boldsymbol{A}_{DS}$. Secondly, we apply the same method as in CKM parameterization \cite{dita_separation_2006}, and transform $\boldsymbol{A}_{DS}$ into a unitary matrix $\boldsymbol{S}^{\prime}$. 

For the first step, as there are three variables $A_1, A_2$, and $A_3$ in $\boldsymbol{E}_S$ and six equations to fulfill for double stochastic matrices, we form the following underdetermined system of equations
\begin{equation} 
\boldsymbol{M}\vec A=\vec R,
\label{Eq:undertermined}
\end{equation}
where
\begin{equation}
    \boldsymbol{M}=\begin{bmatrix}
1 & 0 & NR_{13}^2\\
NR_{21}^2 & 1 & 0\\
0 & NR_{32}^2 & 1\\
NR_{21}^2 & 0 & 1\\
1 & NR_{32}^2 & 0\\
0 & 1 & NR_{13}^2\\
\end{bmatrix},
\end{equation}
\begin{equation}
\vec A= 
\begin{bmatrix}
A_1^2\\A_2^2\\A_3^2
\end{bmatrix},
\vec R =
\begin{bmatrix}
1-R_1^2\\1-R_2^2\\1-R_3^2\\1-R_1^2\\1-R_2^2\\1-R_3^2\\
\end{bmatrix}.
\end{equation}
To minimize the differences $\underset{\vec A} \min  \| \boldsymbol{M}\vec A - \vec R \|$, we adopt the ordinary least square solution 
\begin{equation}
  \vec A_a= (\boldsymbol{M}^{\dag}\boldsymbol{M})^{-1}\boldsymbol{M}^{\dag} \vec R.
\end{equation}
 Once $\vec A_a$ is calculated, we define the energy part of the new matrix $\boldsymbol{E}_S|_{\vec A =\vec A_a}$ as $\boldsymbol{A}_{E}$
\begin{equation}
  \boldsymbol{A}_{E}(i,j):= |\boldsymbol{E}_S|_{\vec A =\vec A_a}(i,j)|^2.
\end{equation}
Due to the underdetermined nature of Eq. \ref{Eq:undertermined}, $\boldsymbol{A}_{E}$ is not in general an exact double stochastic matrix, but is close to one $\boldsymbol{A}_{DS}$. Therefore, we introduce a perturbation matrix $\boldsymbol{\epsilon}$, which satisfies 
$\boldsymbol{A}_{E}-\boldsymbol{\epsilon}  \in \{\boldsymbol{A}_{DS}\}$.  To keep the level of nonreciprocity and reflection small, the perturbation matrix $\boldsymbol{\epsilon}$ should be small enough. As a result, we reshape the goal into a simple optimization problem:
\begin{gather}
  \underset{\epsilon} \min \| \boldsymbol{\epsilon} \|_2 \\
  s.t. \boldsymbol{A}_{E}-\boldsymbol{\epsilon}  \in \{\boldsymbol{A}_{DS} \}.
\end{gather}
Once $\boldsymbol{\epsilon}_{a}$ has been solved, we obtain one $\boldsymbol{A}_{DS}$ corresponding to $\boldsymbol{E}_S$. 

In the second step, we look for a CKM matrix $\boldsymbol{S}_{CKM}$  parameterized by
\begin{equation}
   \begin{bmatrix}
c_{12} & c_{13}s_{12} & s_{12}s_{13}\\
c_{23}s_{12} & -c_{12}c_{13}c_{23}-e^{i\delta}s_{13}s_{23} & -c_{12}c_{23}c_{13}+e^{i\delta}c_{13}c_{23}\\
s_{12}s_{23} & c_{23}s_{13}e^{i\delta}-c_{12}c_{13}s_{23} & -c_{13}c_{23}e^{i\delta}-c_{12}s_{13}s_{23}
\end{bmatrix},
\end{equation}
where $c_{ij}=\cos \theta_{ij}, s_{ij}=\sin \theta_{ij}$,  $\theta_{12},\theta_{13},\theta_{23}$ are angle parameters to be determined, and $\delta$ is a phase parameter. We define the short-hand notations $a:=\sqrt{\boldsymbol{A}_{DS}(1,1)}, b:= \sqrt{\boldsymbol{A}_{DS}(1,2)}, c:= \sqrt{\boldsymbol{A}_{DS}(2,1)},$ and $d:= \sqrt{\boldsymbol{A}_{DS}(2,2)}.
$
Following \cite{dita_separation_2006}, the four parameters are then given by
\begin{gather}
    \cos{\theta_{12}}= a \\
    \cos{\theta_{13}}= \frac{b}{\sqrt{1-a^2}} \\
    \cos{\theta_{23}}= \frac{c}{\sqrt{1-a^2}}
\end{gather}
and 
\begin{multline}
    \cos{\delta}= \frac{-(1-a^2)^2(1-d^2)}{2abc\sqrt{1-a^2-b^2}\sqrt{1-a^2-c^2}}+\\
    \frac{(1-a^2)(b^2+c^2)-b^2c^2(1+a^2)}{2abc\sqrt{1-a^2-b^2}\sqrt{1-a^2-c^2}}.
\end{multline}
In a few unfortunate cases, for which $|\cos \delta| >1$, we approximate the matrix by taking $\delta =0$ ($\cos \delta >1$) or $\pi$ ($\cos \delta <-1$) . In practice, the averaged value of $\| \boldsymbol{\epsilon} \|_2$ is always less than 0.05.
\section{\label{ap:numerical} Numerical RG scheme with replicas for disordered networks }
\begin{figure*}[htbp]
\includegraphics[width=0.9\textwidth]{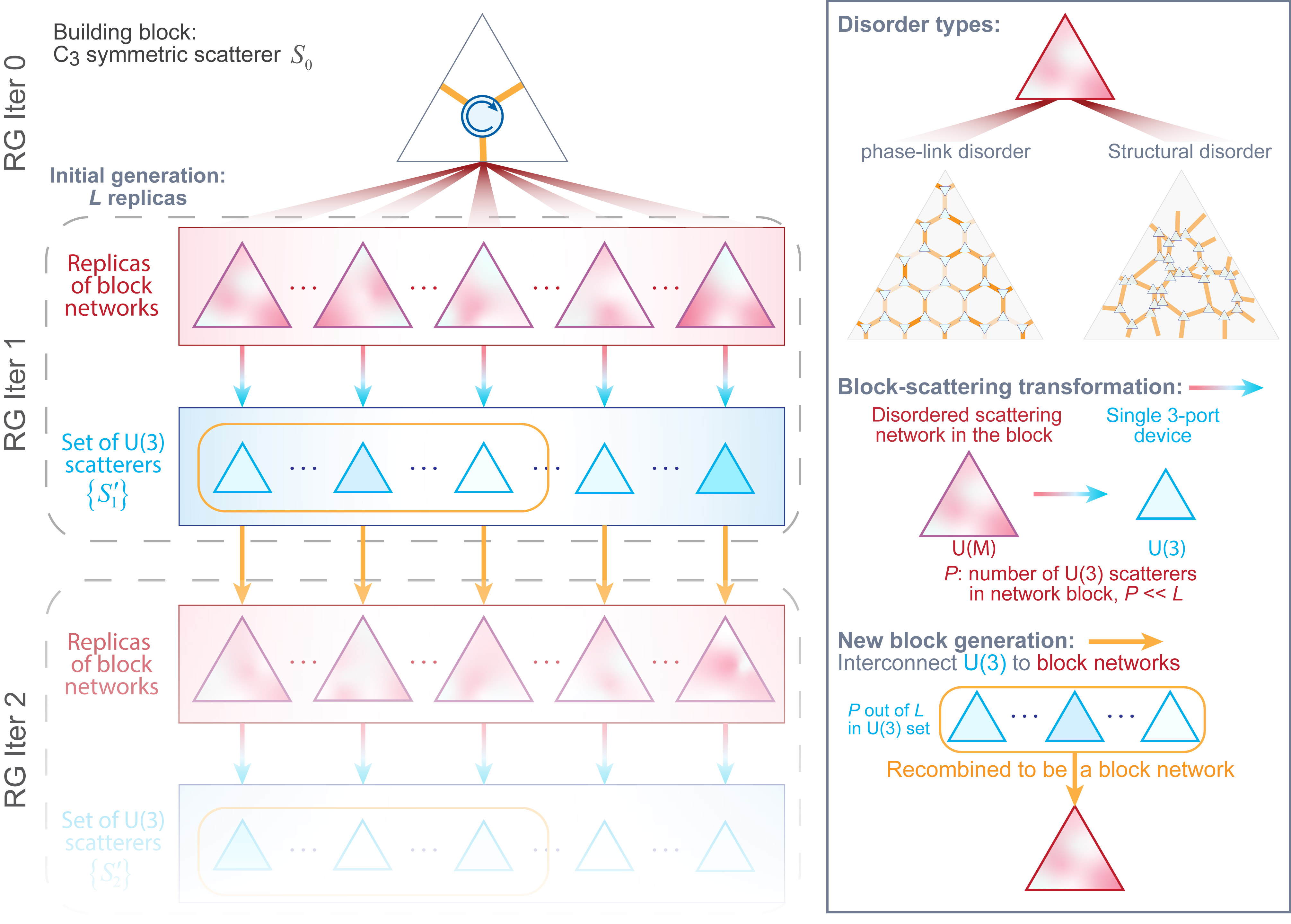}
\caption{\textbf{Numerical RG scheme of disordered scattering networks}. We assume scattering networks subject to disorder (phase-link or structural, left inset) described by specified statistics $P_{disorder}$. To start with, by taking the microscopic three-port scatterer $\boldsymbol{S}_0(\xi,\eta)$ as building blocks, we construct $L$ replicas of triangular block networks (red triangles) for the RG iteration 1. Each replica follows the same network disorder statistics $P_{disorder}$, and contains $P (P \ll L)$ microscopic scatterers. Second, we apply block scattering transformations, which turn each block network into a $U(3)$ scatterer (blue triangles). The scattering matrix set $\bigl\{\boldsymbol{S}_{1}^{\prime}\bigr\}$ of these $U(3)$ scatterers represents the scattering properties of  RG iteration 1. Third, to generate block scattering networks of the iteration 2,  we construct $L$  replicas of triangular block networks in the disorder statistics $P_{disorder}$, and most importantly each replica is composed by $P$ scatterers randomly selected from the set $\bigl\{\boldsymbol{S}_{1}^{\prime}\bigr\}$. By iteratively performing the above three-step process, we obtain the sequence of $\bigl\{\boldsymbol{S}_{n}^{\prime}\bigr\}$, the flow of  probability distributions ($P(T_{CW,n})$, $P(T_{CCW,n})$, and $P(R_{n})$), and the averaged scattering properties $\langle \boldsymbol{S}_n^{\prime} \rangle$.}
\label{fig:numerical RG scheme}
\end{figure*}

The replica scheme employed in our work is detailed in Fig. \ref{fig:numerical RG scheme}. Let us assume that the disorder (phase-link or structural) imparted to the scattering networks conforms to a specified statistical distribution, denoted as $P_{disorder}$. At the first RG iteration, we generate $L$ replicas of triangular block networks, each of which contains $P$ microscopic scatterers described by $\boldsymbol{S}_0(\xi,\eta)$ and generated following the statistics $P_{disorder}$.  Each replica is then transformed into a $U(3)$ scatterer, by applying the block-scattering transformation. The scattering matrix set of these $L$ scatterers forms the $U(3)$ matrix set $\bigl\{\boldsymbol{S}_{1}^{\prime}\bigr\}$. To go to the next iteration, we build $L$ new replicas by randomly drawing $P$ elements of $\bigl\{\boldsymbol{S}_{1}^{\prime}\bigr\}$ and constructing $L$ new block scattering networks following the disorder statistics prescribed by $P_{disorder}$. To make sure the sampling is random enough, we set $L \gg P$.  This procedure is then iterated.

In a sum, what our numerical RG scheme performs is visually represented in the horizontal and vertical exes of Fig. 15. Horizontally, we take the statistic average over replicas, whose number is constant at each iteration. Vertically, we iterate the procedure to approach the thermodynamic limit (under large enough $P$). The number of calculation steps is reduced to a linear function of $n$, expressed as $nL$, as opposed to replica-free schemes like the one of Figs. 5 and 13, which grows as $P^n$. The constant size of the sets $\bigl\{\boldsymbol{S}_{n}^{\prime}\bigr\}$ also allows us to perform consistent statistical analysis as we iterate (Figs. \ref{fig:two examples} and \ref{fig:probab_critical}). This leads to a coherent description of how the scattering properties of disordered network are transformed when networks are scaled up. The numerical RG results shown in this paper assume the settings $L= 4000, P= 100$ and $M_1= M_2= M_3= 3\text{ or }5$.  The only restriction on $P$  is the fact that the block networks indeed have a bulk, namely a network depth of at least $3$, as topological edge transport originates from bulk topology. The selections of $P$ and $L$ only affect the resolution of the boundary. A brief study on the effect of the replica number $L$ is performed around the critical point $\xi=-\eta\approx 0.93$, and shown in Fig. \ref{fig:ap_L_scaling}.



\section{\label{ap:cleanlimit} Topological phases in the clean-limit network}
\begin{figure}[htbp]
\includegraphics[width=0.45\textwidth]{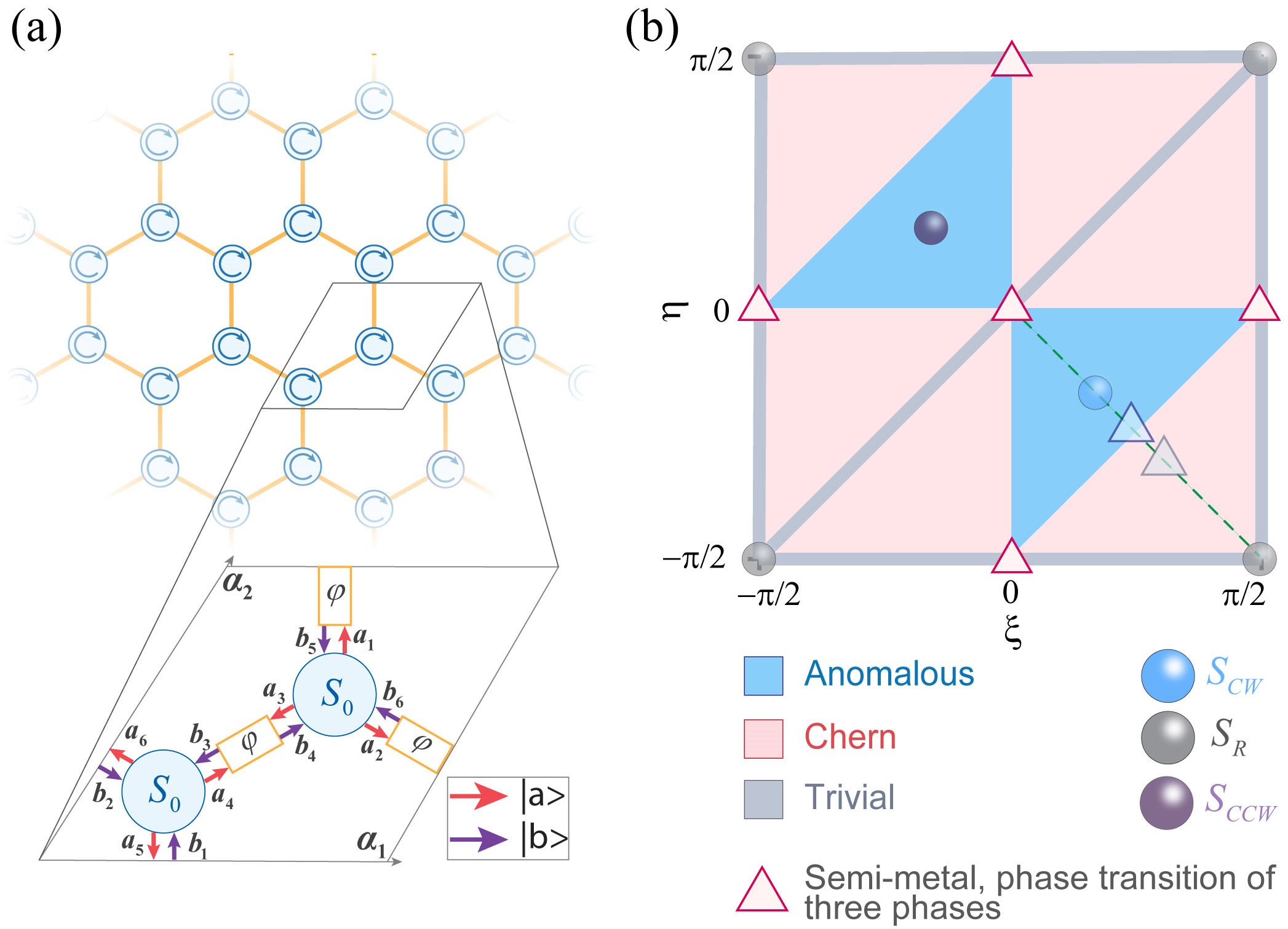}
\caption{\textbf{Clean-limit scattering network and topological phase diagram.} (a) Periodic honeycomb network, with zoomed unit cell. We label the waves propagating in the unit cell, from which an unitary matrix can be derived to describe the scattering process. This unitary matrix in the momentum space leads to an eigen-equation, analogous to the eigen-equation in a Floquet system, with phase delay $\varphi$ taking the role of the quasi-energy.  (b) Topological phase diagram in the $(\xi,\eta)$ plane, showing anomalous, Chern, and trivial phases. The semi- metal cases, marked by red triangles, are the phase transition points of three phases.}
\label{fig:cleanPhasediagram}
\end{figure}
In this part, we review the topological phases and their observables in the clean-limit network. The clean-limit network is defined as a periodic honeycomb network, whose three-port scatterers are identical and all the phase links impart the same phase-delay value $\varphi$, as shown in Fig. \ref{fig:cleanPhasediagram}. Based on the Bloch theorem and the scattering process in the unit cell \cite{pasek_network_2014,delplace_phase_2017,zhang_superior_2021}, we describe the infinite network by a momentum-space eigen-problem involving a unitary scattering matrix $\boldsymbol{S}(\boldsymbol{k})$,
\begin{equation}
    \boldsymbol{S}(\boldsymbol{k}) |b(\boldsymbol{k})\rangle = e^{-i\varphi(\boldsymbol{k})}|b(\boldsymbol{k})\rangle,
\end{equation}
where unitarity leads to a real-valued eigen-phase delay $\varphi$. Plotting the eigen-phases $\varphi (\boldsymbol{k})$ versus momentum $\boldsymbol{k}=(k_x,k_y)$ forms band structures, where the vertical axis is a compact space, namely the circle spanned when $\varphi$ goes through the interval $[0, 2\pi)$.
The above equation evidences the analogy with Floquet band theory defined on unitary time-evolution operators \cite{rudner_band_2020}, allowing for a clear classification of topological phases in scattering networks. Note that $\varphi$ takes the role of the quasi-energy \cite{klesse_spectral_1997,lindner_floquet_2011,rudner_anomalous_2013,liang_optical_2013,delplace_phase_2017,pasek_network_2014,potter_quantum_2020,zhang_superior_2021}. The most important new feature found in unitary or Floquet systems, when compared to topological systems described by a Hermitian eigenvalue problem, is the existence of an extra nontrivial topological phase in the A class, called anomalous. The anomalous phase exhibits topologically protected chiral edge states propagating along system boundaries, albeit they cannot be predicted from the Chern numbers of the bands. Actually, the topology of two-dimensional unitary networks is revealed by the homotopy $\pi_{3}(U(N)) = \mathbb{Z}$, whose elements are the topological gap invariants 
\begin{equation}
W_{gap}(\varphi)=(24\pi^{2})^{-1} \int \textrm{tr}(V^{-1}_{\varphi} \textrm{d} V_{\varphi})^3 \quad \in \mathbb{Z}.
\label{W_gap}
\end{equation} 
In the panel (b) of Fig. \ref{fig:cleanPhasediagram}, with the help of the band structures and topological invariants, we show the topological phase diagram for any possible scatterer $\boldsymbol{S}_0(\xi,\eta)$ in the parameter space of $(\xi,\eta)$.  As shown in Fig. \ref{fig:cleanPhasediagram}, periodic scattering networks can support three phases: anomalous topological phase (blue), Chern phase (pink), and trivial phase (gray). Remarkably, the points corresponding to $\boldsymbol{S}_{CW}$ and $\boldsymbol{S}_{CCW}$ are at the centers of anomalous topological phases, while the trivial phase is centered on $\boldsymbol{S}_R$. In addition, there are special points, the crossing points of three phases, which exhibit semi-metal band structures, represented by red triangles in Fig. \ref{fig:cleanPhasediagram}. 
\begin{figure}[htbp]
\includegraphics[width=0.48\textwidth]{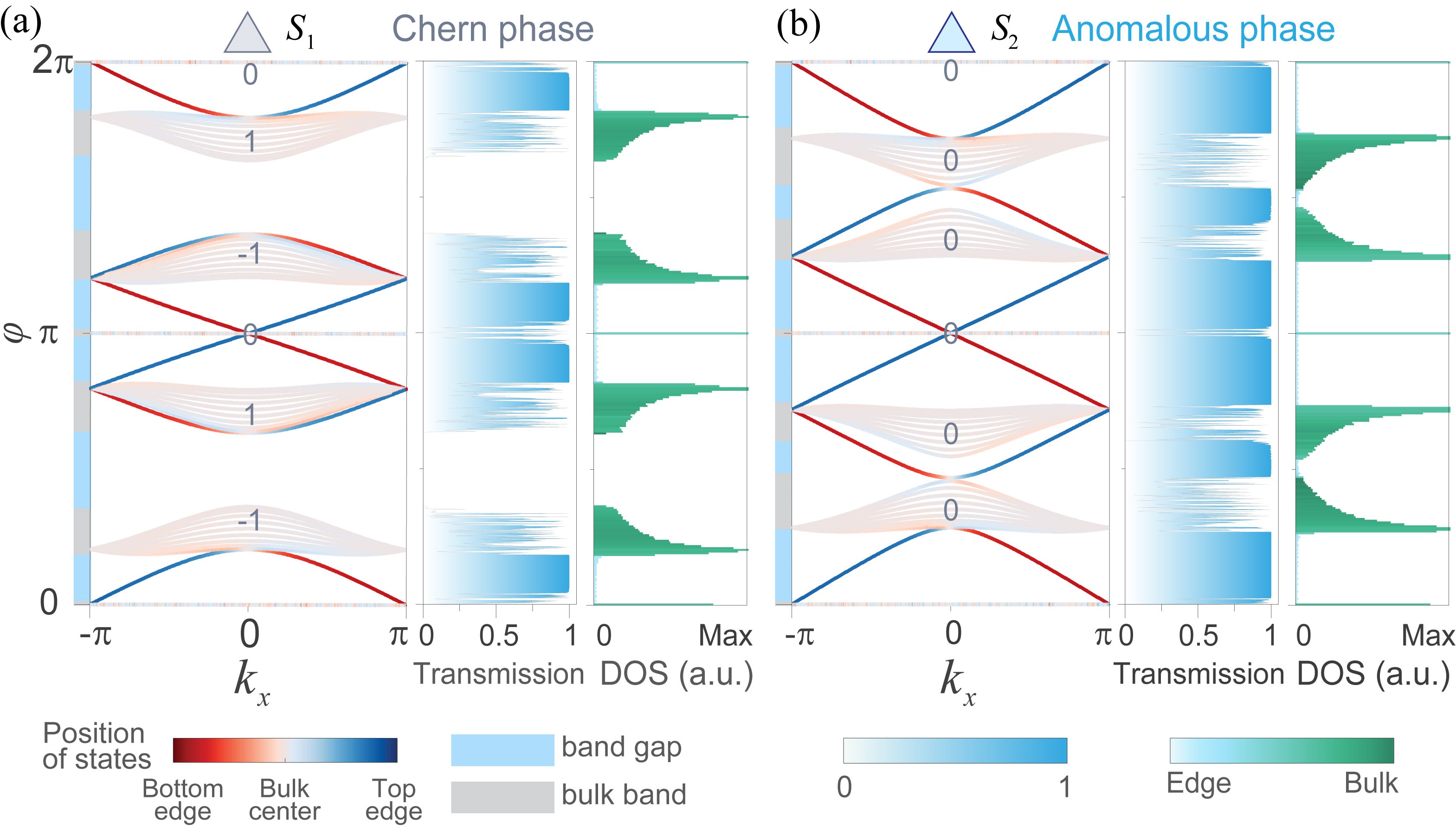}
\caption{\textbf{Topological phases and observables.} Band structures of a supercell, two-port transmissions, density of states (DOS) for the periodic $\boldsymbol{S}_1$ (a) and $\boldsymbol{S}_2$ (b) networks. The numbers on the bands represent the calculated Chern numbers. $\boldsymbol{S}_2$ network is in anomalous phase, which features zero Chern numbers and topological gapless edge states in each band gap. As a contrary,  $\boldsymbol{S}_1$ network is in Chern phase, as demonstrated by the non-zero Chern numbers. In the finite honeycomb networks, we can identify the band structure by checking DOS (rightmost) and two-port transmission (center) versus quasi-energy (namely, phase delay value, $\varphi$).  The unity transmission is mediated by topological edge states with low DOS, while the fluctuating finite transmission along with high DOS indicates bulk states in bands.}
\label{fig:BS_DOS_examples}
\end{figure}
We take $\boldsymbol{S}_1$ and $\boldsymbol{S}_2$ used in the main text as examples, which are marked in Fig. \ref{fig:cleanPhasediagram} as gray and blue triangles, respectively. As shown in the Fig. \ref{fig:BS_DOS_examples} (a), the clean-limit network made of $\boldsymbol{S}_1$ is in the Chern phase, due to the non-zero Chern number of several bands. Its bulk bands in specified ranges of $\varphi$ are consistent with two-port transmissions and high density of states (DOS) evaluated in finite networks, while the topological band gaps are consistent with the unity transmission and the DOS. Its trivial band gap is associated with a blocked transmission and zero DOS. Contrarily, the periodic $\boldsymbol{S}_2$ network is in the anomalous phase, evidenced by the vanishing Chern numbers and the existences of topological edge states in every band gap. There is no trivial gap in the anomalous phase. In a sum, although the periodic $\boldsymbol{S}_1$ network and the periodic $\boldsymbol{S}_2$ network are of distinct topological phases, both networks can support diffusive waves in the bulk, and most importantly both exhibit topological unidirectional edge waves. 

\section{\label{app:subsec} RG criticality around $\xi= -\eta \approx 0.92$}
\begin{figure}[htbp]
\includegraphics[width=0.43\textwidth]{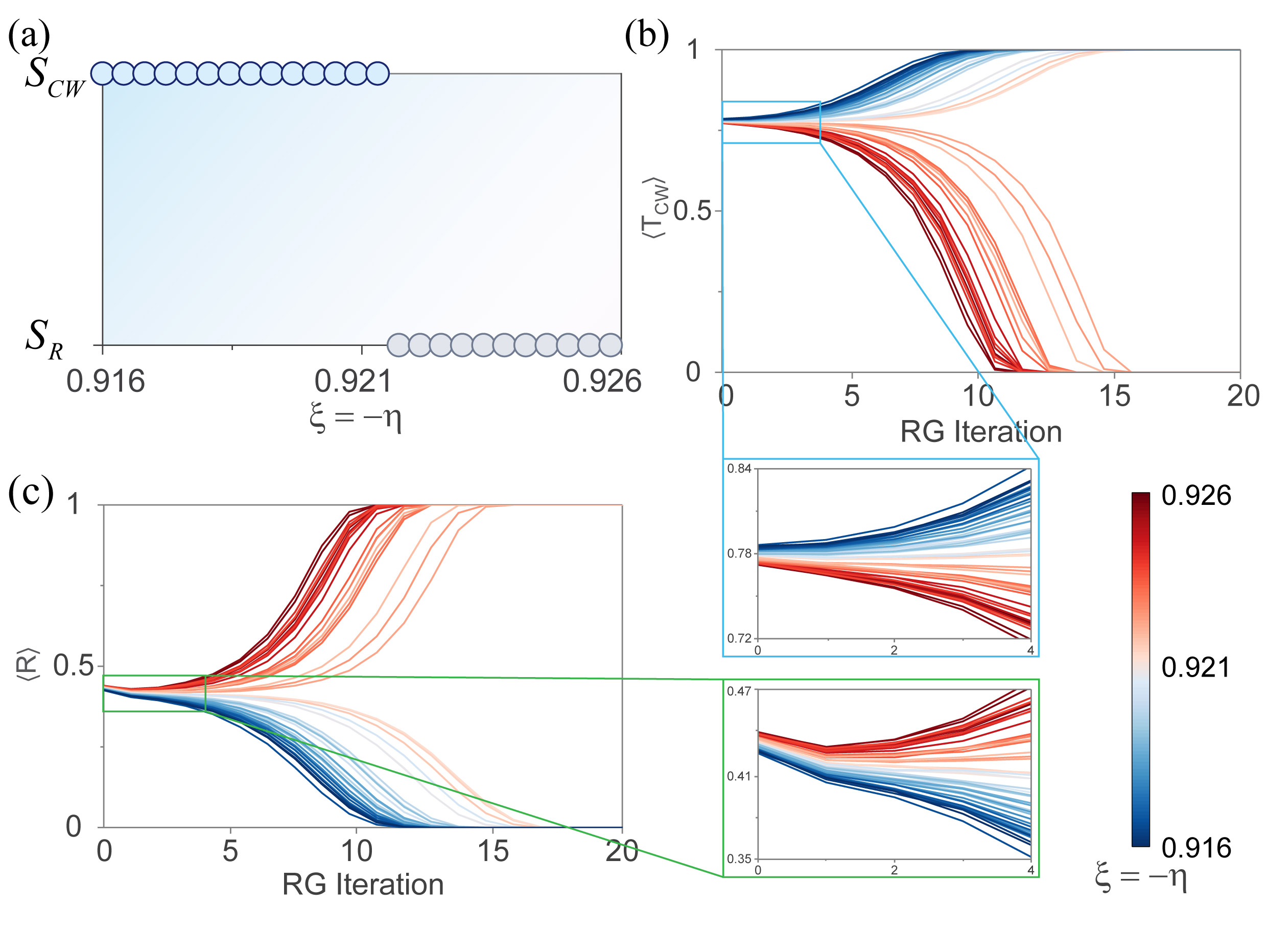}
\caption{\textbf{Evolution of $\langle T_{CW} \rangle$ and $\langle R \rangle$ near the critical point ($\xi=-\eta=0.921$) with increasing renormalization steps.} Panel (a) displays a detailed view of the topological phase transition within the narrow parameter range $\xi=-\eta \in [0.916, 0.926]$, precisely centered at the critical point. Panels (b) and (c) show the flows of $\langle T_{CW} \rangle$ and $\langle R \rangle$, respectively. Initially, the scattering properties of single 3-port scatterers are similar, but as renormalization proceeds iteratively, they diverge, signaling distinct phase transitions. This divergence is characteristic of a saddle point in the RG flow diagram, indicating that the identified phase boundary aligns exactly with the critical point. Replicas used in these RG calculations were $L=8000$ for a clear phase transition. }
\label{figAp:critical behavior}
\end{figure}

This section presents supplementary results that explore the critical renormalization group (RG) behavior around $\xi= -\eta \approx 0.92$. We examine a quite narrow segment in the parameter plane defined by $\xi = -\eta$ within $0.916 < \xi < 0.926$. This range includes the critical point at approximately $\xi = -\eta \approx 0.92$. Three-port scatterers $\boldsymbol{S}_0(\xi,\eta)$ on this segment just differ from each other slightly. However, as shown in Fig. \ref{figAp:critical behavior}(a), the phase-link disordered networks built by these $\boldsymbol{S}_0(\xi,\eta)$ at the left and right parts on this segment converge to opposite attractors: $\boldsymbol{S}_{CW}$ (topological) and $\boldsymbol{S}_R$ (trivial), respectively. To see how the scattering properties of their networks evolve when increasing the network scale, we iteratively perform scattering RG for the disordered scattering networks built by these $\boldsymbol{S}_0(\xi,\eta)$. As the number of renormalization group (RG) iterations increases, corresponding to a scaled-up network size, we check the evolution of the averaged reflection $\langle R \rangle$ and chiral clockwise transmission $\langle T_{CW} \rangle$ defined in Eq. \ref{eq:<S'n>}. $\langle R \rangle$ and $\langle T_{CW} \rangle$ exhibited in Figs. \ref{figAp:critical behavior}(b-c) show that the scattering properties which are nearly identical before performing RG, eventually diverge significantly, aligning with distinct scattering attractors as the number of RG iterations grows. This divergence is indicative of the saddle-point dynamics observed at the critical point in the RG flow diagram. 


\begin{figure}[htbp]
\includegraphics[width=0.48\textwidth]{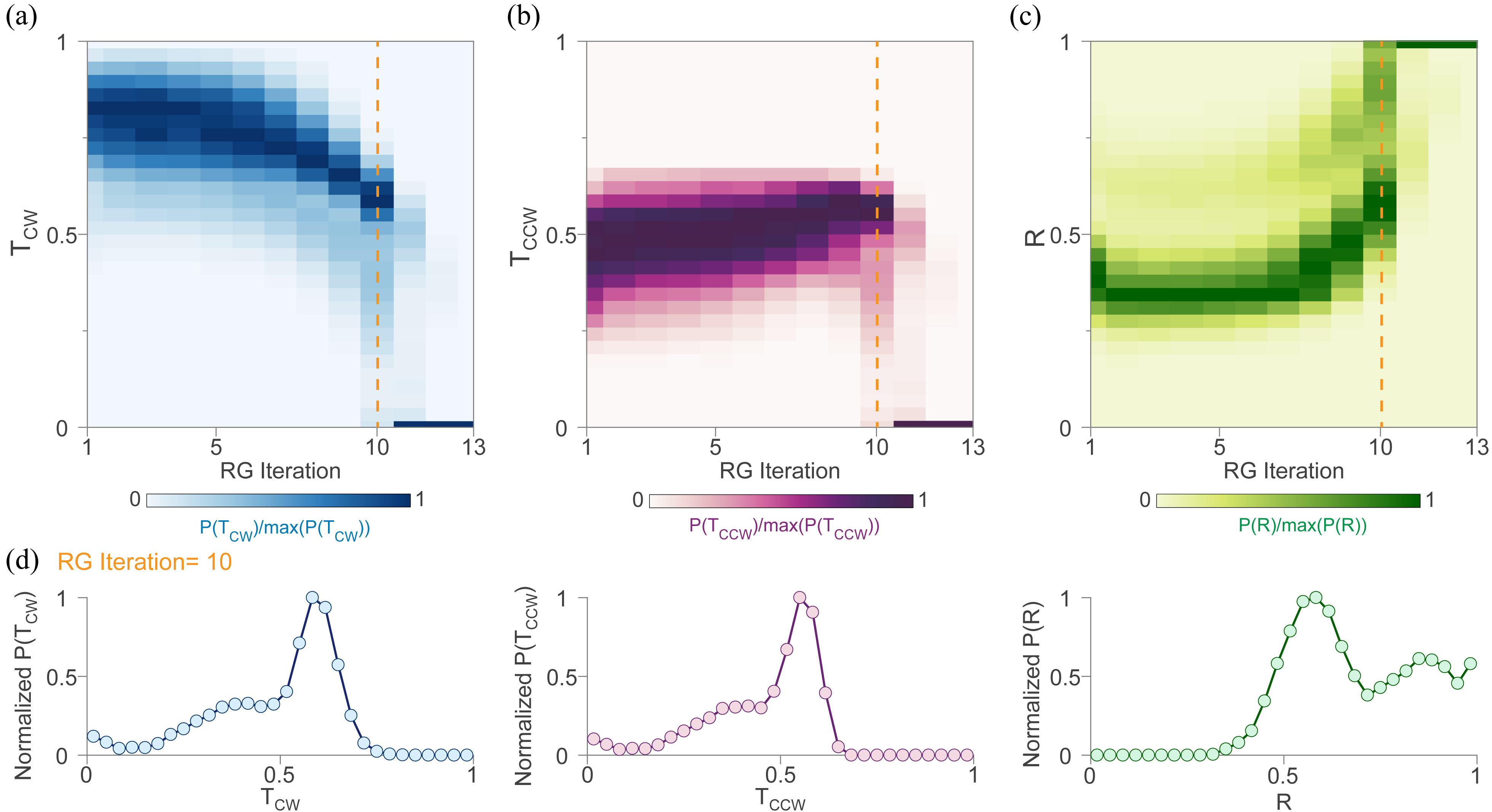}
\caption{\textbf{Collapse to $\boldsymbol{S}_R$ near the critical point $\xi= -\eta= \xi_c \approx 0.921$.} (a) RG transformations of the normalized distributions $P(T_{CW})$, $P(T_{CCW})$, and $P(R)$ of the microscopic scattering matrices at $\xi= -\eta= 0.923$ , namely on the side converging to $\boldsymbol{S}_R$. The small variation of the probability distributions in the first five iterations is characteristic of critical behavior. Dramatically, the distributions change at iteration 10, and converge to the attractor $\boldsymbol{S}_R$. (d) Normalized $P(T_{CW})$, $P(T_{CCW})$, and $P(R)$ at iteration 10. The fact that $P(T_{CW}) \approx P(T_{CCW})$ indicates that the blocks are, on average, reciprocal. This means that networks behave like metals, therefore breaking down quickly under strong disorder.}
\label{figAp:Stat_breaksdown}
\end{figure}

In addition, compared with Fig. \ref{fig:probab_critical} which shows the probability distributions $P(T_{CW})$, $P(T_{CCW})$, and $P(R)$ with the convergence to $\boldsymbol{S}_{CW}$ (left part in Fig.\ref{figAp:critical behavior}(a)), we here provide another RG example to show the collapse to the trivial phase for a point nearby the critical point but on the side of $\boldsymbol{S}_R$. Figs. \ref{figAp:Stat_breaksdown}(a-c) exhibit the evolution of these probability distributions by performing scattering RG for the disordered scattering networks made of $\boldsymbol{S}_0(\xi,\eta)$ at $\xi= -\eta= 0.923$. The slow evolution in the first several iterations indicates the point is around the critical point. Remarkably, chiral transports quickly break down at the $10_{th}$ iteration, and the network at this scale converges to the trivial phase ($\boldsymbol{S}_R$) suddenly. This fast collapse can be explained by the occurrence of ``averaged reciprocal distributions" \cite{wang_anderson_2023}- in which $P(T_{CW}) \approx P(T_{CCW})$ (Fig. \ref{figAp:Stat_breaksdown}(d)). Once arrived there, the system behaves like a metal which quickly get trivially localized under strong disorder.

\begin{figure}[htbp]
\includegraphics[width=0.34\textwidth]{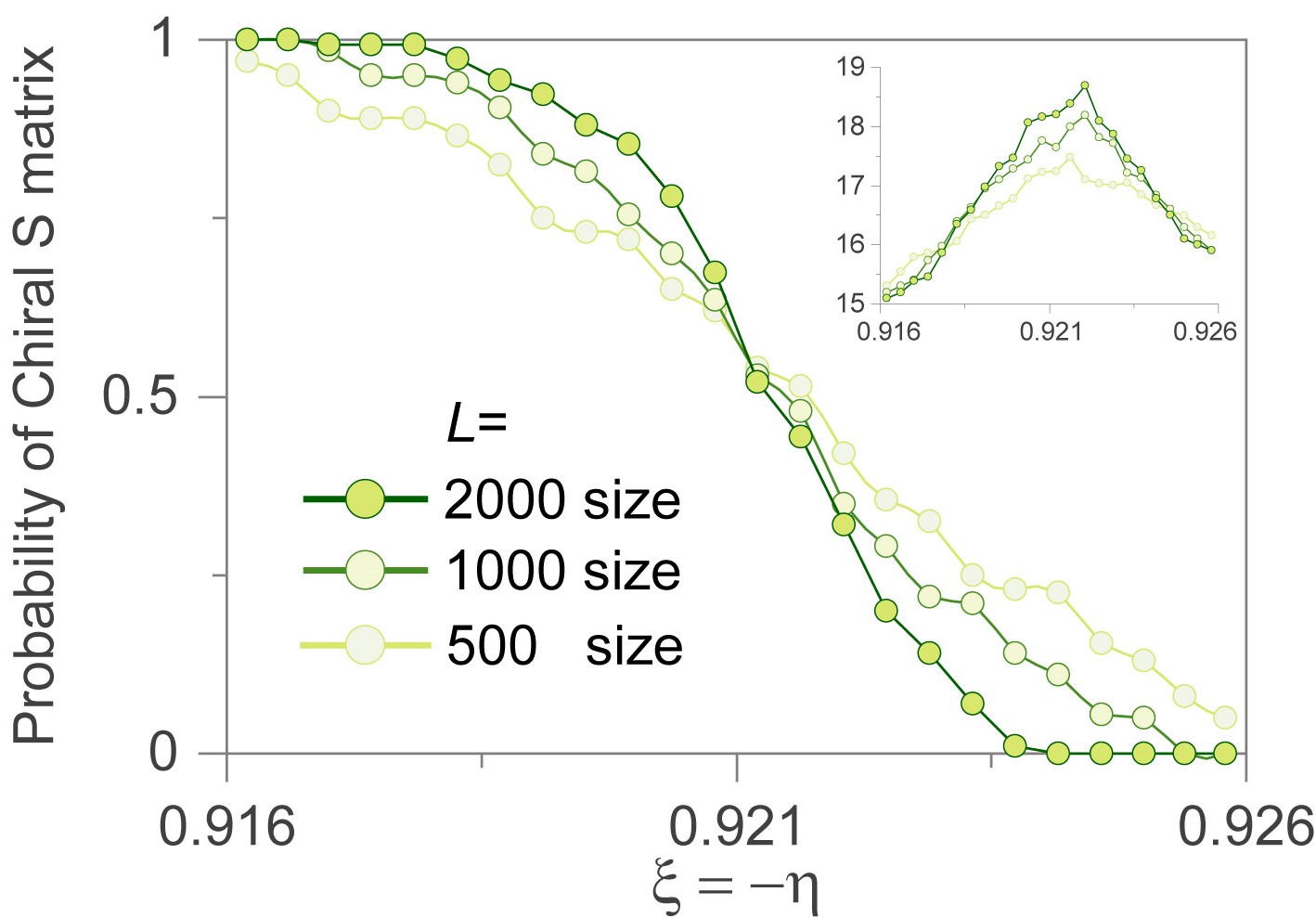}
\caption{\textbf{Effect of increasing the number of replicas $L$ on the probability of scattering attractor around the critical point $\xi= -\eta= \xi_c \approx 0.921$.} With the increasing $L$, the transition from unity probability of $\boldsymbol{S}_{CW}$ to zero becomes sharper, indicating a clearer phase boundary. At the critical point, this probability is independent of $L$.}
\label{fig:ap_L_scaling}
\end{figure}

As discussed in Appendix B, a higher resolution around the critical point requires more replicas (larger $L$) in the numerical RG. This is evidenced in Fig. \ref{fig:ap_L_scaling}.

\section{\label{app:LL} Boundary conditions in localization length calculations and scaling analysis at the other type of critical point: $\xi= -\eta= 0.0947$}

\begin{figure}[!hbtp]
\includegraphics[width=0.48\textwidth]{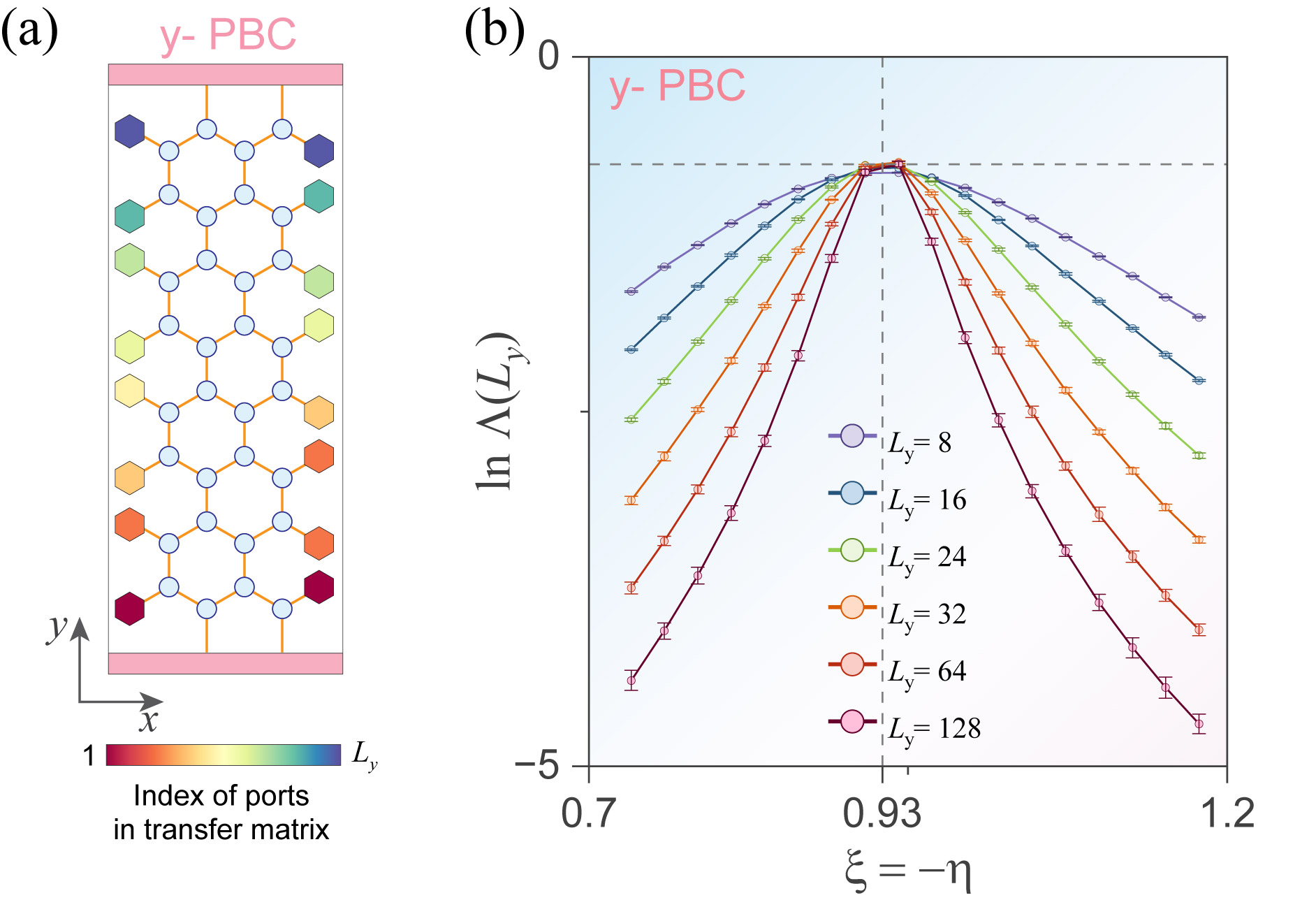}
\caption{\textbf{Scaling of $\Lambda(L_y)$ under y-periodic boundary conditions in the vicinity of the critical point $\xi_c \approx 0.93$ on the line of $\xi= -\eta$.} (a) The periodic boundary condition along the y direction eliminates edge effects, including topological edge states. (b) On both sides of the dashed line, $\Lambda(L_y)$ decreases when increasing the width $L_y$ from 8 to 128, indicating an insulator-insulator transition without revealing the topological nature of one of the insulators.}
\label{fig:PBC_LL}
\end{figure}

When approaching the critical point, especially in the case of insulator- insulator transitions, to reduce the finite size effects, it is better to apply open boundary conditions \cite{yamakage_criticality_2013}. In fact, a study of $\Lambda$ of quasi-1D networks under periodic boundary conditions (PBC) makes it much more difficult to identify a scale-invariant critical point  between the two insulating phases, as shown in Fig. \ref{fig:PBC_LL}. As PBC eliminates edges, topological and trivial insulating phases both show a decrease of the function $\Lambda (L_y)$ when the width $L_y$ is increased, and the peak of this function is relatively flat due to the finite size effect. Therefore, it is not easy to precisely locate the scale-invariant point under PBC, at which the transition between two insulating phases occurs. This is why the main text used open boundary conditions, as in the presence of edges the transition behaves as a metal–insulator crossover, due to the presence of the chiral edge states in the topological insulating networks.

\begin{figure}[!hbtp]
\includegraphics[width=0.48\textwidth]{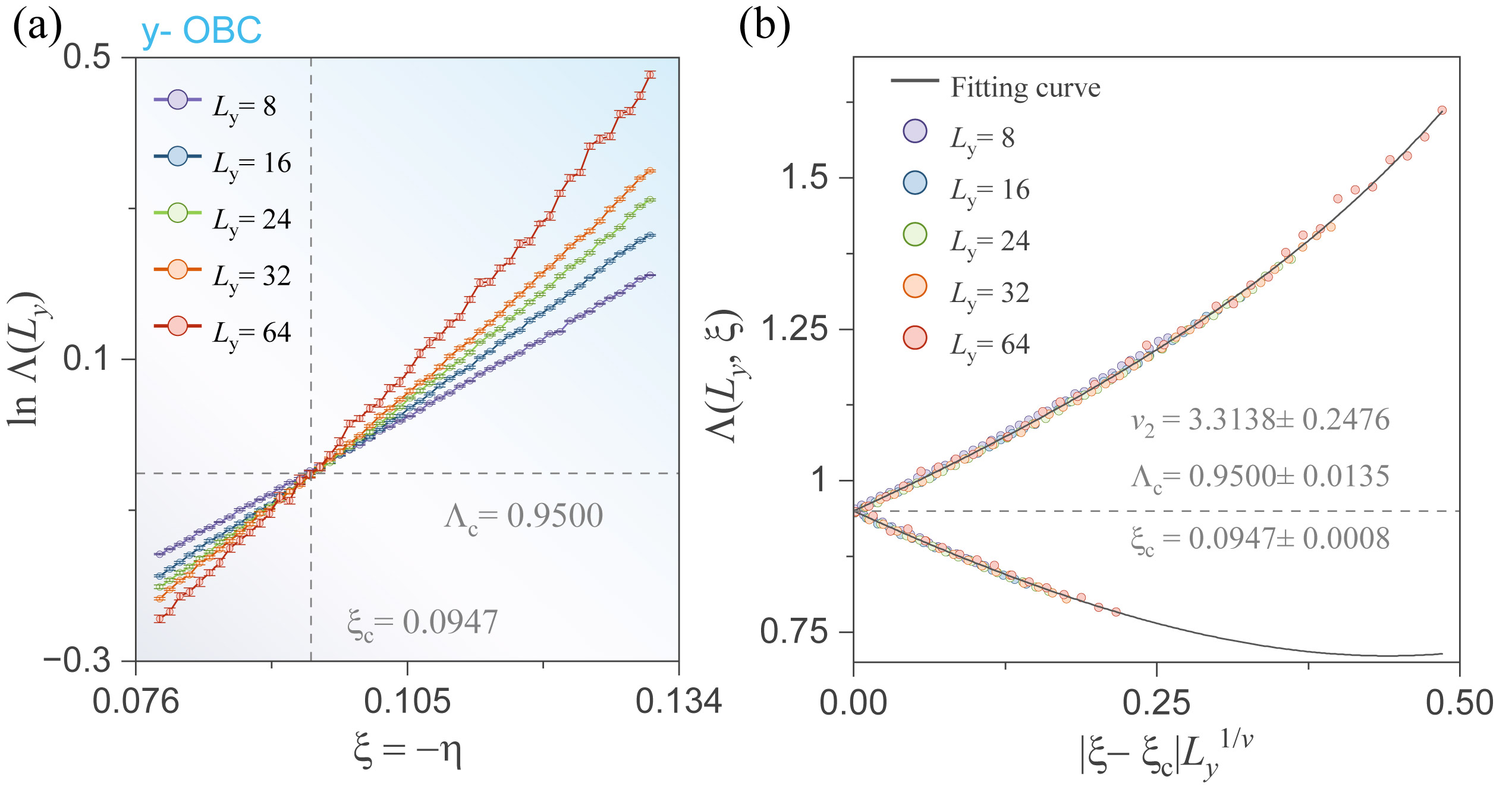}
\caption{\textbf{Scaling analysis in the vicinity of the other type of critical point: $\xi_c \approx 0.0947$ on the line of $\xi= -\eta$.} (a) $\Lambda(L_y)$ at different parameters versus $L_y$, which is increased from $8$ to $64$. (b) Fitted curved and critical parameters with the ansatz of single parameter scaling. Critical exponent $\nu_2$ is estimated to be 3.3138 with critical $\Lambda_c \approx 0.9500$, distinct with the critical exponent $v_1$ in the vicinity of $\xi_c \approx 0.9301$.}
\label{fig:LL_035}
\end{figure}

For a segment defined by $\xi= -\eta \in [0, pi/2]$ in the parameter space, the segment starts from the trivial region, crosses the topological region, and ends in the trivial region, therefore containing two critical points of topological phase transitions. Apart from the one at $\xi_c \approx 0.9301$, we can also turn our view to the other one at $\xi_c \approx 0.0947$. We perform the same scaling analysis as in Fig. 10, and find a distinct critical exponent, $\nu_2\approx 3.3138$.